\documentclass[fleqn,usenatbib]{mnras}

\usepackage{newtxtext,newtxmath}
\usepackage{multirow}
\usepackage{graphicx}	
\usepackage{amsmath}
\usepackage[T1]{fontenc}
\usepackage{adjustbox} 
\usepackage{threeparttable}

\DeclareRobustCommand{\VAN}[3]{#2}
\let\VANthebibliography\thebibliography
\def\thebibliography{\DeclareRobustCommand{\VAN}[3]{##3}\VANthebibliography}

\title[Afterglows from steep jets viewed off axis]{Reverse and forward shock afterglow emission from steep jets viewed off axis}

\author[Ernazar Abdikamalov and Paz Beniamini]{
Ernazar Abdikamalov$^{1,2}$\thanks{ernazar.abdikamalov@nu.edu.kz}
and
Paz Beniamini$^{3,4,5}$\thanks{pazb@openu.ac.il}
\\
$^{1}$Department of Physics, Nazarbayev University, 53 Kabanbay Batyr Ave, Astana 010000, Kazakhstan \\
$^{2}$Energetic Cosmos Laboratory, Nazarbayev University, 53 Kabanbay Batyr Ave, Astana 010000, Kazakhstan \\
$^{3}$Department of Natural Sciences, The Open University of Israel, PO Box 808, Ra’anana 4353701, Israel \\
$^{4}$Astrophysics Research Center of the Open university (ARCO), The Open University of Israel, PO Box 808, Ra’anana 4353701, Israel \\
$^{5}$Department of Physics, The George Washington University, 725 21st Street NW, Washington, DC 20052, USA
}

\date{Accepted XXX. Received YYY; in original form ZZZ}

\pubyear{\the\year{}}

\begin{document}
\label{firstpage}
\pagerange{\pageref{firstpage}--\pageref{lastpage}}
\maketitle

\begin{abstract}
We study the morphology of gamma-ray burst (GRB) afterglows viewed off-axis using a simplified analytical model. We consider steep jets, which are expected to be the most common type. These jets, characterized by steep lateral gradients in energy and Lorentz factor, produce highly beamed emission. The observed signal is dominated by their minimum visible angle at any given time. Consequently, the afterglow morphology depends on when this angle begins to decrease, revealing the inner regions of the jet. Depending on whether this decrease occurs before, at, or after the reverse shock crosses the ejecta, three distinct classes of light curves emerge. In the first scenario, the de-beamed emission can produce a rapidly rising signal even prior to the reverse shock crossing. This is expected in GRBs with long duration, low energy, dense circumburst media, or combinations thereof. In some cases, the ejecta shell can be considered as effectively thick in the inner regions and effectively thin in the outer regions. For forward shocks, the temporal slopes in both regimes are identical, which makes it hard to detect the transition. Reverse shocks, however, have distinct temporal slopes, allowing potential detection of the transition in light curves if their emission surpasses that of the forward shock. The characteristic synchrotron frequency of de-beamed emission evolves independently of jet structure for forward shocks but depends on the lateral energy and Lorentz factor gradients for the reverse shock, with slower evolution for steep energy and shallow Lorentz factor gradients.
\end{abstract}


\section{Introduction}
\label{sec:intro}

As the GRB jet interacts with the circumstellar medium, two shocks are formed: the forward shock (FS), which propagates into the external medium, heating and accelerating it, and the reverse shock (RS), which moves backward into the ejecta, heating and decelerating it \citep[e.g.,][]{sari95hydrodynamics, Meszaros97Optical}. The electrons accelerated by these shocks produce the afterglow emission \citep[e.g.,][]{sari98spectra, Sari99Predictions, Kobayashi00Light, Zhang06Physical, Peer06Signature}. These afterglows, observable from X-ray to radio wavelengths, evolve over timescales from seconds to years, providing probes of jet dynamics, energetics, and surrounding environments \citep{Frail01Beaming, Li12Comprehensive, Zhao20Statistical}.

The evolution of the ejecta depends on whether the RS becomes relativistic before traversing the ejecta or remains sub-relativistic \citep{sari95hydrodynamics}. The former occurs in `thick shells', while the latter takes place in `thin shells' \citep{Zhang22semi}. In thick shells, the RS becomes relativistic and significantly decelerates the ejecta. By the time the RS crosses the shell, most of the ejecta kinetic energy has been transferred to the shocked external medium \citep{sari97hydrodynamics}. In contrast, for thin shells, the RS remains sub-relativistic, resulting in negligible deceleration of the ejecta. However, as the RS crosses the shell, the FS sweeps up sufficient material to transfer most of the ejecta initial kinetic energy to the external medium \citep[e.g.,][]{kobayashi1999hydrodynamics}. Following shock crossing, the ejecta transitions into a decelerating self-similar expansion phase \citep[e.g.,][]{Piran93Hydrodynamics}.

For a jet with Lorentz factor $\Gamma$, relativistic beaming confines emission to a cone with a half-angle of $\Gamma^{-1}$ \citep[e.g.,][]{Sari99Jets, Woods99Constraints, Yamazaki02Xray}. Observers can detect emission only if they are within the beaming cone of the emitting region \citep{granot02offaxis}. Consequently, although GRBs are randomly oriented \citep[e.g.,][]{Rhoads97How}, we are likely to observe jets that are on-axis or slightly off-axis \citep{Nakar02Detectability, Beniamini19Observational}. As the jet decelerates, however, the beaming angle widens, revealing further regions on longer time scales \citep{Granot05AfterglowObservations, Granot05Afterglow}. Studying these events is important for understanding the full GRB population and jet structures \citep{Huang04Rebrightening, Lamb21Inclination, Duque22Flares, Li24Nature}. For a recent review, see \citet{Salafia22Structure}.

The detection of gravitational waves from the neutron star merger GW170817, along with the subsequent GRB, provides strong evidence for an off-axis viewing scenario \citep{abbott2017gw170817b, Granot17Lessons, Lazzati17Off, Lamb17Electromagnetic, Gill18Afterglow, Kathirgamaraju18Off}. Initially, off-axis viewing resulted in faint early emission due to relativistic beaming, but as the jet decelerated and its emission spread, radiation from the inner, more luminous regions became observable \citep{Mooley18Superluminal, Ghirlanda19Compact, Margutti18Binary, Troja18outflow, Hajela19Two}. Future simultaneous electromagnetic and gravitational wave observations \citep[e.g.,][]{Beniamini2019,Keinan24Coordinated} will enable more detailed studies.

There have been numerous efforts to study the off-axis light curves and angular structure of jets through analytical methods \citep[e.g.,][]{Rossi02Afterglow, Granot03Constraining, Kumar03Evolution, Panaitescu03Effect, Rossi04polarization, Granot05AfterglowObservations, Eichler06Case, beniamini20afterglow, Beniamini22Robust} and numerical modeling \citep[e.g.,][]{DeColle12Gamma_1, Alexander18Decline, Granot18Off, Lamb18GRB, Lazzati18Late, Xie18Numerical, Ryan20afterglowpy, Gottlieb223DGRMHD}. Despite significant progress, important gaps remain. First, in thick-shell GRBs, the jet begins decelerating as soon as the RS becomes relativistic, even before the RS crosses the ejecta. This deceleration can be significant, especially for GRBs with very long durations, low energies, or those in dense circumburst environments (or combinations of these factors). The observational signatures of this phase have not been explored thoroughly. Additionally, while recent studies advances our understanding of off-axis RS emission \citep{Fraija19Short, Lamb19Reverse, Gill23GRB, Pang24Reverse_1, Pang24Reverse_2}, the full parameter space remains uncharted. This work aims to bridge these gaps. 

While numerical modeling offers detailed insights, it is constrained by the significant computational cost of exploring vast parameter spaces and the degeneracies among intrinsic parameters, which can obscure their individual contributions to light curves. In contrast, the simplicity of analytical modeling enables more efficient exploration of the parameter space while offering deeper insights into the underlying physical dependencies. In our work, we adopt the latter approach. 

We study the morphology of off-axis light curves from steep jets. Such jets, which are expected to be common, are characterized by steep gradients in energy and Lorentz factor, where emission from smaller viewing angles dominates over that from larger angles. Consequently, the observed signal primarily originates from the minimum visible angle, a key assumption of our model. We study light curves from FS and RS and explore how they depend on the jet parameters and viewing angle. The morphology depends on when the minimum visible angle decreases. This timing, relative to the reverse shock crossing, creates three light curve classes. If the decrease occurs early, a rapidly rising signal can appear before the shock crossing. Some ejecta may behave as a thick shell in the inner regions and a thin shell in the wings. We find that while FS light curves exhibit identical temporal slopes, RS light curves display distinct slopes between thick and thin shells. This distinction suggests that the transition may be detectable in light curves, provided the RS emission rises above that of the FS.

This paper is organized as follows: Section \ref{sec:method} outlines our methodology, Section \ref{sec:results} presents the results, and Section \ref{sec:conclusion} provides conclusion.

\section{Method}
\label{sec:method}

\subsection{Jet dynamics}
\label{sec:thin_vs_thick}

We assume that the initial kinetic energy per unit solid angle, $\epsilon$, and Lorentz factor, $\Gamma_0$, are given by smoothly broken power-law functions:
\begin{equation}
\label{eq:jet_structure}
\frac{\epsilon(\theta)}{\epsilon_\mathrm{c}} = \Theta^{-a}, \quad \frac{\Gamma_0(\theta) - 1}{\Gamma_{\mathrm{c},0} - 1} = \Theta^{-b}, \quad \Theta \equiv \sqrt{1 + \left(\frac{\theta}{\theta_\mathrm{c}}\right)^2},
\end{equation}
where $\theta$ is the angle from the jet’s axis and $\theta_\mathrm{c}$ is the half-opening angle of the jet core. The key advantage of this simple prescription is that it enables independent variation of the angular profiles of energy and Lorentz factor, which allows us to investigate their individual effects on their off-axis signal. We assume that the circumstellar medium density is described by $\rho = AR^{-k}$. 

For thin shells, where the RS remains sub-relativistic, the ejecta experiences little deceleration prior to shock crossing. As a result, the Lorentz factor at shock crossing, $\Gamma_\Delta(\theta)$, remains equal to the initial Lorentz factor, $\Gamma_0(\theta)$ \citep{Yi13early}. Deceleration begins only after the RS traverses the shell, which takes place at \citep{beniamini20afterglow}
\begin{equation}
\label{eq:tdec}
    t_\Delta^\mathrm{tn}(\theta) = \left[ \frac{(3-k) \epsilon(\theta)}{ 2^{3-k} c^{5-k} A \Gamma_0 (\theta)^{2(4-k)}} \right]^{\frac{1}{3-k}} = t_{\mathrm{c},\Delta}^\mathrm{tn}
    \Theta^{\frac{2(4-k)b-a}{3-k}},
\end{equation}
where $t_{\mathrm{c},\Delta}^\mathrm{tn}$ is the shock crossing time for the jet core. Unlike thin shells, the shock crossing time for thick shells, which we denote  $t_{\Delta}^\mathrm{tk}$, is independent of angle. Due to this reason, $t_{\mathrm{c},\Delta}^\mathrm{tk} = t_\Delta^\mathrm{tk}\approx \Delta_0/c$ for thick shells, where $\Delta_0$ is the initial width of the shell. To distinguish between thin and thick shells, the notation $t_\Delta^\mathrm{tn}(\theta)$ will denote the shock crossing time for thin shells, whereas $t_\Delta^\mathrm{tk}$ will represent the shock crossing time for thick shells. The notations $t_\Delta$ and $t_\mathrm{c,\Delta}$ will represent contexts applicable to both cases.

The thick shells begin decelerating before shock crossing when the RS becomes relativistic at
\begin{equation}
\label{eq:tN}
t_\mathrm{N} (\theta) = t_\Delta^\mathrm{tk} \left[ \frac{\Gamma_\Delta (\theta) }{\Gamma_0 (\theta) } \right]^{\frac{2(4-k)}{2-k}} = t_\mathrm{\Delta}^\mathrm{tk}
\left[ \frac{\Gamma_{\mathrm{c},\Delta}} {\Gamma_{\mathrm{c},0}} \right]^{\frac{2(4-k)}{2-k}} \Theta^{\frac{2(4-k)b-a}{2-k}},
\end{equation}
where 
\begin{equation}
\label{eq:gamma_delta}
\Gamma_\Delta(\theta) = 
\left[ \frac{l(\theta)}{\Delta_0} \right]^\frac{3-k}{2(4-k)} =
     \left[ \frac{l_\mathrm{c}}{\Delta_0} \right]^\frac{3-k}{2(4-k)} \Theta^{\frac{-a}{2(4-k)}} =
    \Gamma_\mathrm{c, \Delta} \Theta^{\frac{-a}{2(4-k)}}.
\end{equation}
$\Gamma_{\Delta}(\theta)$ represents the critical Lorentz factor that marks the transition between the thin-shell regime ($\Gamma_0 < \Gamma_{\Delta}$) and the thick-shell regime ($\Gamma_0 > \Gamma_{\Delta}$). $l(\theta)$ is the angle-dependent Sedov length,
\begin{equation}
    l(\theta) = \left[ \frac{(3-k) \epsilon(\theta) }{ 4 \pi A m_\mathrm{p} c^2  }  \right]^{1/(3-k)} =
    l_\mathrm{c} \Theta^\frac{-a}{3-k},
\end{equation}
where $l_\mathrm{c}$ is the Sedov length for the jet core. 

For some parameters, the ejecta can exhibit thick-shell behavior in the inner regions and thin-shell behavior in the outer regions \citep{Pang24Reverse_2}. The thick shell approximation is valid for $\Gamma_0(\theta) > \Gamma_\Delta(\theta)$. This condition is violated at angle
\begin{equation}
\label{eq:theta_tr}
    \theta_\mathrm{tr} \approx \theta_\mathrm{c} \left( 
    \frac{\Gamma_{\mathrm{c},0}}{\Gamma_{\mathrm{c},\Delta}}
    \right)^\frac{2(4-k)}{2(4-k)b-a},
\end{equation}
which represents the boundary between thin and thick shells. 

Assuming adiabatic evolution without lateral spreading, the LF before and after shock crossing can be expressed as 
\begin{equation}
\label{eq:gamma_vs_t}
\Gamma(\theta,t) = \Gamma_{\mathrm{c},\Delta} \Theta^{-b_\mathrm{p}} \left(\frac{t}{t_\mathrm{c,\Delta}} \right)^{-g}.
\end{equation}
The values of $g$ and $b_\mathrm{p}$ before and after shock crossing are summarized in Table~\ref{tab:g_bp_values}. Notably, for all cases except the thin shell RS, $b_\mathrm{p} = a / 2(4-k)$, meaning it depends solely on $a$, which governs the angular energy profile, and is independent of $b$, the parameter controlling the angular profile of the initial Lorentz factor. Instead, for the thin shell RS, $b_\mathrm{p}$ depends on both $a$ and $b$,
\begin{equation}
    b_\mathrm{p} = \frac{a(7-2k)-2b(4-k)}{4(k^2-7k+12)},
\end{equation}
a distinction with important implications, as we discuss below.

We assume that the post-shock-crossing evolution of the jet follows the \citet{blandford76fluid} solution. While it provides an excellent approximation for thick shells \citep{kobayashi1999hydrodynamics}, for thin-shell shocked ejecta, where the RS is sub-relativistic, it is not strictly valid. However, it still provides a good estimate for the temporal evolution of the Lorentz factor \citep{kobayashi00optical}, which is the key factor governing off-axis emission. Since our work focuses on the qualitative aspects of light curves, the accuracy of this solution is sufficient for our purposes. Furthermore, this solution is the only one available for any value of $k$, which allows us to perform a detailed parameter study. 

The above solution holds in both the thin and thick shell limits but may be less accurate in the intermediate regime \citep[e.g.,][]{Zhang22semi}. We assume a uniform radial jet structure at all angles, which may not apply to jets with complex morphology \citep[e.g.,][]{Gottlieb223DGRMHD} or those powered by late-time central engine activity \citep[e.g.,][]{Sari00Impulsive}. In such cases, or when magnetic fields significantly affect jet dynamics \citep[e.g.,][]{Zhang05Gamma}, the deceleration time and RS crossing time may not coincide in the thin shell limit, unlike in our simplified model. These effects may alter the off-axis afterglow light curve in non-trivial ways. A detailed study of these effects is left for future work.

\subsection{De-beamed emission}
\label{sec:db_emission}

For an observer at $\theta_\mathrm{obs}$, the smallest visible angle $\theta_\mathrm{min}$ can be obtained from the equation 
\begin{equation}
\label{eq:theta_min}
    \theta_\mathrm{obs} - \theta_\mathrm{min} = \Gamma(\theta_\mathrm{min}, t)^{-1}.
\end{equation}
For $\theta_\mathrm{min} \gg \theta_\mathrm{c}$, Eq.~(\ref{eq:gamma_vs_t}) can be written as
\begin{equation}
\label{eq:gamma_theta_min}
 \Gamma(\theta_\mathrm{min}, t) \propto \theta_\mathrm{min}(t)^{-b_\mathrm{p}} \, t^{-g}. 
\end{equation}
For $\theta_\mathrm{min}(t) \ll \theta_\mathrm{obs}$, $\theta_\mathrm{obs} - \theta_\mathrm{min}(t)  \approx \theta_\mathrm{obs} $, so Eq~(\ref{eq:gamma_theta_min}) leads to $\Gamma(\theta_\mathrm{min}, t) = \mathrm{const}$. Combining this with Eq.~(\ref{eq:gamma_theta_min}), we obtain \citep{Gill18Afterglow}
\begin{equation}
\label{eq:thetamin_vs_t}
    \theta_\mathrm{min}(t) \propto t^{-{g}/{b_\mathrm{p}}}.
\end{equation}
The angle $\theta_\mathrm{min}(t)$ increases with time for $b_\mathrm{p}>0$, which is the case for thick shells (for both FS and RS) and for thin-shell FSs. However, for thin-shell RSs, $b_\mathrm{p}$ is negative for $b>a(7-2k)/2(4-k)$, so $\theta_\mathrm{min}(t)$ given by Eq.~(\ref{eq:thetamin_vs_t}) increases with time. This is an artifact of the approximations used to derive Eq.~(\ref{eq:thetamin_vs_t}) and it does not occur in the numerical solution of Eq.~(\ref{eq:theta_min}). Furthermore, observational data do not support this parameter space \citep{Beniamini19Observational}. Therefore, we exclude it from our analysis.

To calculate de-beamed emission, we first derive the expressions for the flux by separating the dependencies on $\theta$ and $t$ \citep{Gill18Afterglow}. For the commonly observed spectral regime $\nu_\mathrm{m} < \nu < \nu_\mathrm{c}$, the flux is given by \citep{sari98spectra, granot02shape}\footnote{The expressions for other spectral regimes can be obtained using similar method as that applied to the $\nu_\mathrm{m} < \nu < \nu_\mathrm{c}$ regime, so, for clarity, we present the calculation process only for this case.}
\begin{equation}
    F_\nu \propto f_\Omega F_{\nu,\text{max}} \nu_\mathrm{m}^{(p-1)/2},
\end{equation}
where $p$ is the power-law index for the distribution of accelerated electrons and $f_\Omega \sim \theta^2 \Gamma(\theta,t)^2 $ is the beaming correction factor \citep{Gill18Afterglow}.

We now express $F_{\nu,\text{max}}$ and $\nu_\mathrm{m}$ in terms of energy, initial Lorentz factor, shock crossing time, and the observer time. These dependencies are provided in \citet{Yi13early} and summarized in Table~\ref{tab:dep_on_E_LF} for completeness. Subsequently, we express energy as $\propto \theta^{-a}$ and the initial Lorentz factor as $ \propto \theta^{-b}$, leading to an expression for $F_\nu$ in the form
\begin{equation}
\label{eq:F_nu_theta_t}
    F_\nu \propto \theta^\zeta t^\tau,
\end{equation}
where $\zeta$ and $\tau$ can depend on $a$, $b$, $k$, and $p$. The condition $\zeta = 0$ separates steep jets ($\zeta < 0$), where the emission is dominated by smaller angles, from shallow jets ($\zeta > 0$), where the emission is dominated by larger angles \citep{Beniamini22Robust}. In our work, we consider steep jets. 

To obtain de-beamed emission, we replace $\theta$ in Eq.~(\ref{eq:F_nu_theta_t}) with $\theta_\mathrm{min}(t)$ given by Eq.~(\ref{eq:thetamin_vs_t}), which results in an expression of the form $F_\nu \propto t^\alpha$. The expressions for the temporal slopes $\alpha$ are provided in Table~\ref{tab:temporal_indices} for the FS and RS, before and after shock crossing, as well as for arbitrary $k$ and $p$. This method, validated by comparisons with numerical integration over the jet emissivity \citep{beniamini20afterglow,Beniamini22Robust}, accurately captures the qualitative features of the off-axis light curve, making it well suited for our study.
  
\section{Results}
\label{sec:results}

\begin{figure}
\centering
\includegraphics[width=0.99 \linewidth]{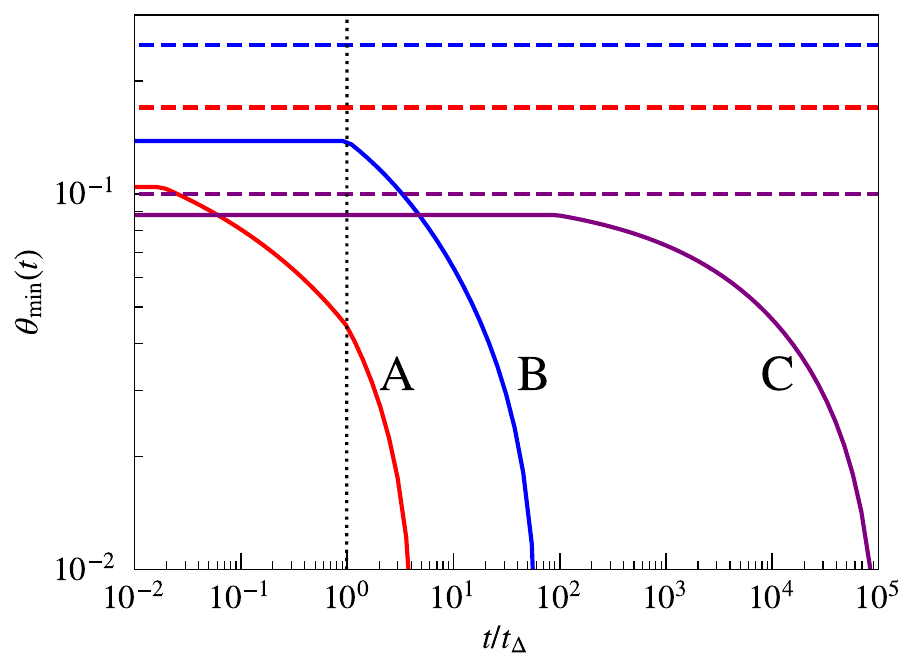}
\caption{Evolution of $\theta_\mathrm{min}(t)$ for cases A, B, and C. In case A, $\theta_\mathrm{min}(t)$ decreases before shock crossing. This occurs in thick shells where the jet decelerates pre-shock. In case B, $\theta_\mathrm{min}(t)$ decreases at shock crossing, which occurs for thin shells where deceleration starts at shock crossing. Case C features $\theta_\mathrm{min}$ decreasing after shock crossing. This happens in a strongly beamed region, so de-beaming occurs well after shock crossing. The dashed lines show $\theta_\mathrm{obs}$. This is possible in both thin and thick shells. In this example, the initial Lorentz factors is $\Gamma_\mathrm{c,0}=200$ for cases A and B. Case A is a thick shell, while case B is a thin shell. Case C is a thin shell with $\Gamma_\mathrm{c,0}=800$. All these cases correspond to FSs. The evolution of $\theta_\mathrm{min}(t)$ for RS exhibits qualitatively similar trends.} 
\label{fig:theta_min}
\end{figure}

\begin{figure*}
\centering
\includegraphics[width=0.99 \linewidth]{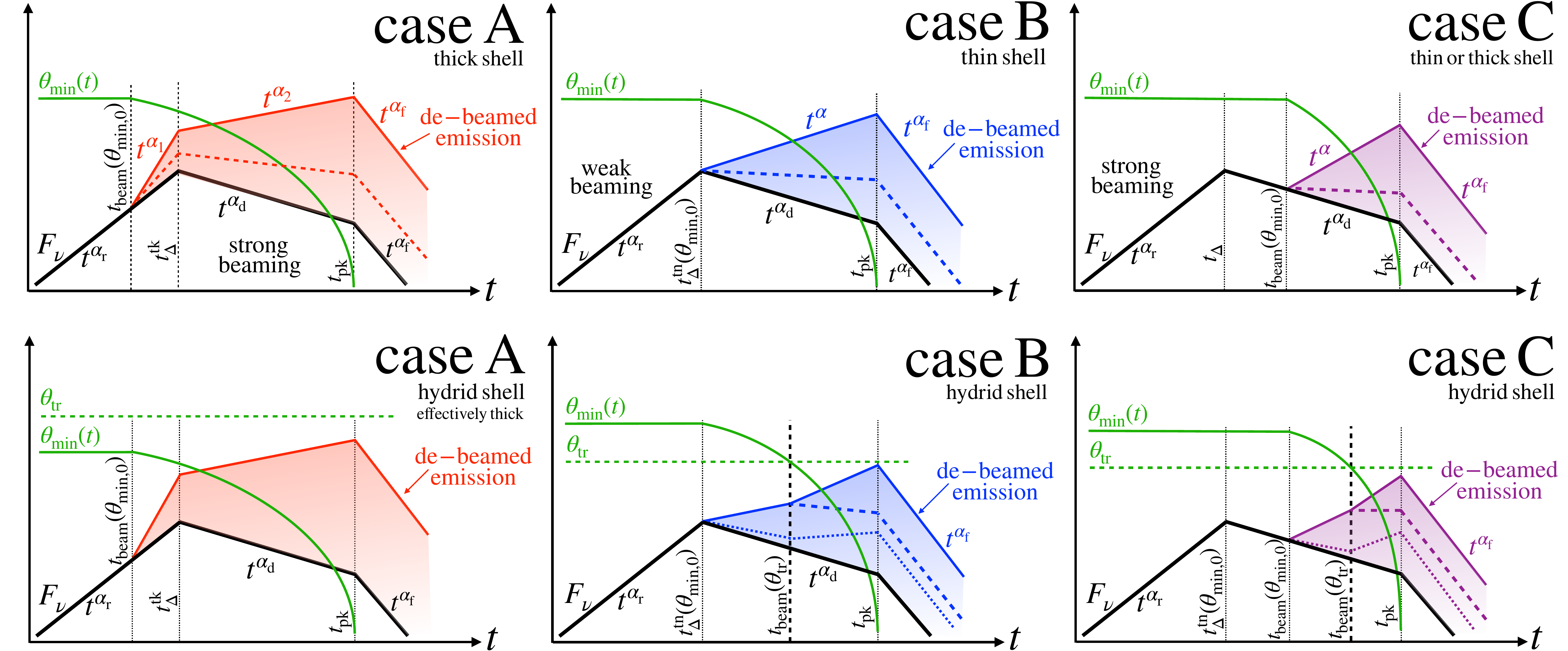}
\caption{Schematic depiction of the distinct morphologies of the light curves from de-beamed emission, corresponding to the three cases of $\theta_\mathrm{min}$ evolution shown in Fig.~\ref{fig:theta_min}. The solid black lines show the light curves for the "on-axis" case for a jet with energy and Lorentz factor at the initial minimum visible angle, $\theta_\mathrm{min,0}$. The upper panels correspond to jets that are purely thin or thick, while the bottom panels depict hybrid jets with thick-shell cores and thin-shell wings. The green lines show the evolution of $\theta_\mathrm{min}(t)$. The red (left panel), blue (center panel), and purple (right panel) lines depict the de-beamed emission for cases A, B, and C. Depending on the jet steepness, the de-beamed emission may exhibit a lower temporal slope, as illustrated by dashed lines. These lines can generally pass through any point in the shaded region. The vertical dashed black lines indicate the transition time from the thin-shell to the thick-shell region.} 
\label{fig:LC_cases}
\end{figure*} 

\subsection{General overview}
\label{sec:qualitative}

If emission is dominated by material at the minimum visible angle $\theta_\mathrm{min}(t)$, the light curve morphology is governed by the temporal evolution of this angle. For $\theta_\mathrm{min}(t)$ to decrease and thus expose the inner and more luminous regions of the jet, two conditions must hold: (1) the jet must decelerate, and (2) the beaming angle $\Gamma (\theta_\mathrm{min},t)^{-1}$ must exceed $\theta_\mathrm{min}(t)$. If the second condition is not met, emission from $\theta_\mathrm{min}(t)$ remains collimated, and the observer sees emission mostly from $\theta_\mathrm{obs} \approx \theta_\mathrm{min}(t) \approx \theta_\mathrm{min,0}$. 

A jet can be classified as either a thick shell, a thin shell, or a combination of both. In the latter case, the inner regions ($\theta < \theta_\mathrm{tr}$; cf. Eq.~\ref{eq:theta_tr}) fall within the thick shell regime, while the outer regions ($\theta > \theta_\mathrm{tr}$) belong to the thin shell regime. We refer to such configurations as hybrid jets. Below, we first discuss scenarios where the jet is characterized exclusively as a thin or thick shell. Subsequently, in Section~\ref{sec:thin2thick}, we explore the hybrid jets.

Consider a jet that can be characterized as either thick or thin shell. The evolution of $\theta_\mathrm{min}(t)$ can be categorized into three cases based on when it begins to decrease: before, at, or after the RS crosses the ejecta, referred to as cases A, B, and C, respectively. In Case A, $\theta_\mathrm{min}(t)$ decreases before shock crossing. This is possible only in thick shells, as thin shells cannot decelerate before the RS crossing \citep[e.g.,][]{sari95hydrodynamics}. Case B occurs in thin shells, where $\theta_\mathrm{min}(t)$ starts decreasing when the RS crosses the ejecta. The observer must be located in the weakly beamed region, where $\Gamma_0 (\theta_\mathrm{min,0})^{-1} \sim \theta_\mathrm{min,0}$. Case C occurs in the strongly beamed region, $\Gamma_0 (\theta_\mathrm{min,0})^{-1} \ll \theta_\mathrm{min,0}$, where de-beaming happens well after the shock crossing as the jet slows sufficiently. This can occur in both thin and thick shells. Fig.~\ref{fig:theta_min} shows three examples of $\theta_\mathrm{min}(t)$ evolution for these three cases. 

The light curves for these three cases are shown schematically in the upper panels of Fig.~\ref{fig:LC_cases}. The solid black lines represent the light curve for the `on-axis' case, corresponding to a jet with energy and Lorentz factor equal to those at the initial minimum visible angle, $\theta_\mathrm{min,0}$. Before shock crossing, it rises, peaking at the shock crossing time $t_\Delta$, and then declining as $\propto t^{\alpha_\mathrm{d}}$ \citep[e.g.,][]{Gao15Morphological}\footnote{For certain parameters, the peak of the on-axis light curve may occur before shock crossing \citep[e.g.,][]{Gao13complete} However, the manner in which de-beamed emission contributes to the light curve remains unaffected by the specific shape of the on-axis light curve. Thus, for simplicity, we focus on the case where the on-axis light curve peaks at the moment of shock crossing.}. At the jet break time, an `on-axis' jet slows down enough for its entire structure to become visible to the observer. The flux steepens by $ \propto \Gamma^2 \propto t^{-2g}$, i.e., $\alpha_\mathrm{f} = \alpha_\mathrm{d}-2g$, where the values of $g$ are listed in Table \ref{tab:g_bp_values} \citep[e.g.,][]{granot02offaxis, Panaitescu07Jet}. This discussion assumes that the structure of the jet remains fixed with time. In reality, it is expected that $t_{\rm jb}$ is also the approximate time at which the jet can start significantly spreading sideways \citep[e.g.,][]{Sari99Jets}. This slows down the blastwave faster than with no spreading and generally results in a more steeply declining lightcurve after this point \citep{Lamb21GRB}. As the degree and speed at which jet spreading occurs is still a topic of active investigation \citep[e.g.,][]{DeColle12Simulations}, we present here the results for the case of no spreading. A generalization to a spreading case is straightforward, once the dynamics are determined. 

In Case A (top left panel, Fig.~\ref{fig:LC_cases}), the observer initially sees highly collimated emission, mostly along its line of sight, so $\theta_\mathrm{obs} \approx \theta_\mathrm{min,0}$. Since $\theta_\mathrm{min}(t)$ (green line) decreases before the shock crossing, the inner regions of the jet become visible early, leading to a faster rise in the light curve, as shown by the red line. This happens when the Lorentz factor at $\theta_\mathrm{min,0}$ drops below $1/\theta_\mathrm{min,0}$. This time can be estimated using Eq.~(\ref{eq:gamma_vs_t}),
\begin{equation}
\label{eq:t_db1}
    t_\mathrm{beam} (\theta_\mathrm{min,0}) \approx t_\mathrm{c,\Delta} \frac{\left(\theta_\mathrm{obs} \Gamma_\mathrm{c,\Delta} \right)^{1/g}}{(\theta_\mathrm{obs}/\theta_\mathrm{c})^{b_\mathrm{p}/g}} .
\end{equation}
After shock crossing, while on-axis emission declines, de-beamed emission may continue rising as additional inner regions become visible, as shown by the solid red line. The peak is reached when the jet core become visible\footnote{In Fig.~\ref{fig:LC_cases}, the peak time, $t_\mathrm{pk}$, for an off-axis observer and the jet break time for an on-axis observer are shown to coincide. However, these times do not necessarily align and depend on the jet structure and on $\theta_\mathrm{obs}$. For example, in a `top-hat' jet model, if the off-axis observer is positioned at $\theta_\mathrm{obs} > \theta_\mathrm{c}$, the peak $t_\mathrm{pk}$ occurs $\sim (\theta_\mathrm{obs}/\theta_\mathrm{c})^{1/g}$ times later than the jet break for the on-axis observer.}. We can estimate this time using Eq.~(\ref{eq:gamma_vs_t}) and the conditions $\theta_\mathrm{min}(t) \rightarrow \theta_\mathrm{c}$ and $\Gamma(\theta_\mathrm{min}, t) \approx 1 / \theta_\mathrm{obs}$, which leads to
\begin{equation}
\label{eq:t_pk}
    t_\mathrm{pk} \approx t_\mathrm{c,\Delta} \frac{
    \left(\theta_\mathrm{obs} \Gamma_\mathrm{c,\Delta}
    \right)^{1/g} }{2^{b_\mathrm{p}/2g}}.
\end{equation}
For some parameters, the de-beamed emission after shock crossing may decline with time, albeit at a slower rate than the on-axis emission. In this case, the light curve peaks at $t_\Delta^\mathrm{tk}$ and experiences a change of slope at $t_\mathrm{pk}$, as shown with the dashed red lines. After the jet core becomes visible, the flux decays as $\propto t^{\alpha_\mathrm{f}}$, similar to post-jet-break flux for on-axis observers \citep{beniamini20afterglow}. Note that the ratio of timescales $t_\mathrm{beam}$ and $t_\mathrm{pk}$, which can be identified from observations, could provide useful constraints on the relationship between $\theta_\mathrm{obs}$, $\theta_\mathrm{c}$, $\Gamma_\mathrm{c,\Delta}$, and $b_\mathrm{p}$, where the latter generally depends on $k$, $a$, and, in the case of thin-shell RS, $b$ (cf. Table~\ref{tab:g_bp_values}). 

Case A is expected when the RS significantly decelerates the ejecta, which leads to a  small Lorentz factor at shock crossing $\Gamma_\Delta$. This can occur in GRBs with an very long duration, low energy, a dense circum-burst medium, or any combination of these factors (cf. Eq.~\ref{eq:gamma_delta}). 

In Case B (top center panel of Fig.~\ref{fig:LC_cases}), $\theta_\mathrm{min}(t)$ (green line) begins to decrease at the moment of shock crossing, initiating the de-beamed emission (blue lines). The initial minimal visible angle $\theta_\mathrm{min,0}$ can be obtained from the condition $\Gamma_0^{-1}(\theta_\mathrm{min,0}) = \theta_\mathrm{obs} - \theta_\mathrm{min,0} \approx \theta_\mathrm{obs}$,
\begin{equation}
\label{eq:theta_min0}
\theta_\mathrm{min,0} \approx \theta_\mathrm{c} \left( \theta_\mathrm{obs} \Gamma_\mathrm{c,0} \right)^{1/b}.  
\end{equation}
At this angle, the RS crosses the ejecta at a time
\begin{equation}
    \label{eq:t_db2}
    t_\Delta^\mathrm{tn} (\theta_\mathrm{min,0}) \approx t_{\mathrm{c},\Delta}^\mathrm{tn} \left( \theta_\mathrm{obs} \Gamma_\mathrm{c,0} \right)^\frac{2(4-k)b-a}{(3-k)b},
\end{equation}
which can be obtained by substituting Eq. (\ref{eq:theta_min0}) into (\ref{eq:tdec}). If de-beamed emission is rising ($\alpha>0$), similar to case A, the peak is reached when the jet core becomes visible, which happens at 
\begin{equation}
\label{eq:t_pk_B}
    t_\mathrm{pk} \approx t_{\mathrm{c},\Delta}^\mathrm{tn} \frac{
    \left(\theta_\mathrm{obs} \Gamma_\mathrm{c,0}
    \right)^{1/g} }{2^{b_\mathrm{p}/2g}}.
\end{equation}
Otherwise, the peak happens at $t_\Delta^\mathrm{tn} (\theta_\mathrm{min,0})$ and the light curve experience a change of slope at $t_\mathrm{pk}$, as shown by dashed blue lines. Similar to case A, the ratio of timescales $t_\Delta^\mathrm{tn} (\theta_\mathrm{min,0})$ and $t_\mathrm{pk}$ could provide useful relationships between $\theta_\mathrm{obs}$, $\Gamma_\mathrm{c,0}$, and $b_\mathrm{p}$. 

In Case C (top right panel, Fig.~\ref{fig:LC_cases}), the observer must initially be located in the strongly beamed regime, so $\theta_\mathrm{min}(t)$ decreases well after the shock crossing, when the beaming angle $\Gamma(\theta_\mathrm{min},t)^{-1}$ exceeds $\theta_\mathrm{min}(t)$. This delay uncovers the inner regions later, resulting in postponed de-beamed emission. If this emission is rising ($\alpha>0$), then the light curve has two peaks, as illustrated by the purple lines. Otherwise, only one peak is present, as shown by the dashed purple lines. The first peak happens when the shock crosses the ejecta. The dip between the two peaks happens when the material below $\theta_\mathrm{obs} \approx \theta_\mathrm{min,0}$ becomes visible, i.e., when $\Gamma(\theta_\mathrm{obs}, {t}_\mathrm{dip})^{-1} \approx \theta_\mathrm{obs}$, which leads to Eq.~(\ref{eq:t_db1}). The second peak is reached when the jet core becomes visible, i.e. at $t_\mathrm{pk}$ given by Eq.~(\ref{eq:t_pk}). 

The ratio of two timescales $t_\mathrm{pk}$ and $t_\mathrm{beam}(\theta_\mathrm{min,0})$, which can be obtained from observations,
\begin{equation}
    \frac{t_\mathrm{pk}}{t_\mathrm{beam}} \approx 
    \left( \frac{\theta_\mathrm{obs} }{ \sqrt{2} \theta_\mathrm{c}} \right)^{b_\mathrm{p}/g},
\end{equation}
allows us to put a constraint on relationship between $\theta_\mathrm{obs}$, $\theta_\mathrm{c}$, and $b_\mathrm{p}$ \citep{beniamini20afterglow}. The latter depends on $k$, $a$, and (in the case of thin-shell RS) on $b$  (cf. Table~\ref{tab:g_bp_values}). The value of $k$ can be obtained from the temporal slope before and after the shock crossing prior to the commencement of the de-beamed emission \citep[e.g.,][]{Tian22Constraining}.

While cases B and C have been previously identified (e.g., as cases 1B and 1A, respectively, by \citealt{beniamini20afterglow}), case A, to the best of our knowledge, has not yet been discussed in the literature.

\subsection{Thin-to-thick transition}
\label{sec:thin2thick}

As noted above, for some parameters, the jet core may lie in the thick-shell regime ($\theta < \theta_\mathrm{tr}$, cf. Eq.~\ref{eq:theta_tr}), while the outer regions correspond to the thin shell regime ($\theta > \theta_\mathrm{tr}$). To observe the transition, the observer must initially see the thin shell region ($\theta_\mathrm{min,0} >  \theta_\mathrm{tr}$). As the jet decelerates, $\theta_\mathrm{min}(t)$ migrates towards the inner regions of the jet. The thick shell part becomes visible when $\theta_\mathrm{min}(t)$ reaches the transition angle $\theta_\mathrm{tr}$. This can happen in cases B and C, but not in A. To understand why, we analyze each case individually below.

The bottom-left panel of Fig.~\ref{fig:LC_cases} shows a schematic plot of $\theta_\mathrm{min}$ as a function of time for case A. Since this case occurs in thick shells, the minimum visible angle, $\theta_\mathrm{min,0}$, must fall within the thick shell region, ensuring that the transition angle $\theta_\mathrm{tr}$ exceeds $\theta_\mathrm{min,0}$. Consequently, the emission from $\theta \ge \theta_\mathrm{tr}$ is weaker than that from $\theta_\mathrm{min}(t)$ and remains buried under it. As the jet decelerates, the beaming effect weakens, revealing the inner, brighter regions of the jet and initiating de-beamed emission at $t_\mathrm{beam} (\theta_\mathrm{min,0})$. 
The corresponding light curve is shown with red lines in the bottom-left panel of Fig.~\ref{fig:LC_cases}. At $t_\Delta^\mathrm{tk}$, the shock crosses the thick shell and continues traversing the thin shell. However, as noted earlier, the emission from this region is overshadowed by the emission from the inner regions, rendering it invisible to the observer. 

The bottom-center panel of Fig.~\ref{fig:LC_cases} shows $\theta_\mathrm{min}$ as a function of time for case $\mathrm{B}$. Unlike case A, the observer initially sees the thin shell region, i.e., $\theta_\mathrm{min,0} > \theta_\mathrm{tr}$. The RS crosses the thick shell ejecta, after which it continues traversing the thin shell wing. Eventually, the RS shock crosses the shell at $\theta_\mathrm{min,0}$, causing the material at this angle to decelerate and forcing $\theta_\mathrm{min}$ to decrease. This exposes inner regions of the jet, resulting in de-beamed emission. The corresponding light curve is shown with blue lines in the bottom-center panel. When $\theta_\mathrm{min}(t)$ reaches $\theta_\mathrm{tr}$, the thick shell part becomes visible. This happens when $\Gamma(\theta,t)\theta$ at $\theta_\mathrm{tr}$ falls below 1, enabling us to estimate this time using Eq.~(\ref{eq:gamma_vs_t}), 
\begin{equation}
\label{eq:t_db_tr}
    t_\mathrm{beam} (\theta_\mathrm{tr}) 
    \approx 
    t_\mathrm{c,\Delta}^\mathrm{tk} \frac{\left(\theta_\mathrm{obs} \Gamma_\mathrm{c,\Delta} \right)^{1/g}}{(\theta_\mathrm{tr}/\theta_\mathrm{c})^{b_\mathrm{p}/g}}.
\end{equation}
This time is shown with the vertical dashed black line.
The jet core becomes visible at $t_\mathrm{pk}$ given by Eq.~(\ref{eq:t_pk}). The ratio of the two timescales
\begin{equation}
\label{eq:ratio_t_tr}
    \frac{t_\mathrm{pk}}{t_\mathrm{beam}} \approx 
    \left( \frac{\theta_\mathrm{tr} }{ \sqrt{2} \theta_\mathrm{c}} \right)^{b_\mathrm{p}/g},
\end{equation}
provides a constraint on the relationship between $\theta_\mathrm{tr}$, $\theta_\mathrm{c}$, and $b_\mathrm{p}$. 

In case $\mathrm{C}$, depicted in the bottom-right panel of Fig.~\ref{fig:LC_cases}, the situation is similar to B but with a key distinction: due to the high degree of beaming at the initial $\theta_\mathrm{min,0}$, the de-beamed emission begins well after the shock crossing, when $\Gamma(\theta_\mathrm{min}(t))\theta_\mathrm{min}(t)$ falls below 1 (shown with purple lines). The material at $\theta_\mathrm{tr}$ becomes visible even later, once $\Gamma(\theta)\theta$ at that angle also drops below 1. The transition time, given by Eq.~(\ref{eq:t_db_tr}), is indicated by the vertical dashed black lines. The ratio of peak times (\ref{eq:ratio_t_tr}) applies to case C too. 

For FSs, as we will see below, the temporal slopes in both the thick and thin shell regimes are identical, which make such transitions difficult to detect observationally using temporal slopes alone. In contrast, RSs exhibit distinct temporal slopes for thick and thin shells across most $a$ and $b$ values, suggesting the potential to observe this transition in RS light curves. This scenario is shown with dashed blue and purple lines in the bottom center and right panels of Fig.~\ref{fig:LC_cases} for cases B and C, respectively. Moreover, since thick shell RS becomes relativistic, its emission should exceed that of the thin-shell RS \cite[e.g.,][]{Pang24Reverse_2}. Therefore, the thin-to-thick transition may appear as a hump in the light curve, provided that it can rise above the emission from the FS \citep{gao15reverse}.

\subsection{Forward shock}
\label{sec:fs_emission}

\begin{figure}
\centering
\includegraphics[width=0.99 \linewidth]{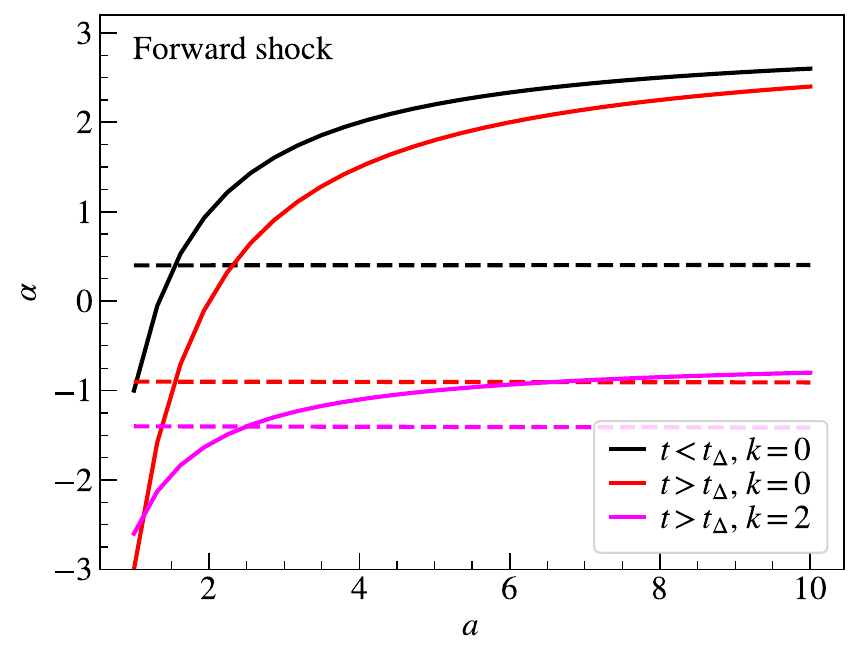}
\caption{Temporal slope of de-beamed emission (solid lines) from FS as a function of $a$, the parameter that governs the jet angular energy profile, for the spectral regime $\nu_\mathrm{m} < \nu < \nu_\mathrm{c}$. For comparison, the temporal slope for an on-axis observer is shown with dashed lines. Black lines indicate emission before shock crossing (relevant for thick shells only), while red and magenta lines represent post-shock-crossing emission for $k=0$ and $k=2$, respectively (relevant for both thick and thin shells). $\alpha$ increases with $a$, reflecting stronger core emission in steeper jets. Unlike the RS, $\alpha$ for FS is identical for thin and thick shells and is independent of $b$, the parameter that governs the initial Lorentz factor angular profile.} 
\label{fig:FS}
\end{figure}

For forward shocks, temporal slope $\alpha$ increases with $a$ in all cases. This is due to the steeper energy distribution along $\theta$ for larger $a$ (cf. Eq.~\ref{eq:jet_structure}), resulting in a steeper rise in de-beamed emission. Moreover, $\alpha$ has the same value for thin and thick shells and it is independent of $b$, the parameter governing the angular profile of the initial Lorentz factor. These trends hold in all spectral regimes (analytical expressions for $\alpha$ for various spectral regimes are provided in Table~\ref{tab:temporal_indices}). 

As an example, the value of $\alpha$ for the spectral regime $\nu_\mathrm{m} < \nu < \nu_\mathrm{c}$ before shock crossing is
\begin{equation}
\label{eq:alpha_FS_b}
    \alpha = \frac{2(k-2)}{a}-\frac{1}{4}k(p+5)+3,
\end{equation}
which is shown with solid black lines in Fig.~\ref{fig:FS} as a function of $a$ for $k=0$. As mentioned before, de-beamed emission before shock crossing can happen only in case A, i.e., for thick shells only. For $a \gtrsim 2$, $\alpha$ surpasses the temporal slope of the on-axis emission (represented by the black dashed line), signifying that the jet becomes sufficiently steep so that de-beamed emission from inner regions dominates the emission from $\theta_\mathrm{min,0}$. For large $a$, $\alpha$ converges to $3-k(p+5)/4$\footnote{For large $a$, the jet can be approximated as a `top-hat' jet. In such a situation, the light curve is dominated by strongly de-beamed emission from the core. This can lead to rise steeper than $\propto t^3$ \citep[see Appendix A of][]{Beniamini23Swift}.}. Hence, FS emission can produce a rapid rise in the afterglow even before shock crossing. As noted above, this behavior is expected in GRBs with very long durations, low energies, dense circumburst environments, or a combination of these factors. 

The circumstellar density slope, $k$, significantly influences the value of $\alpha$: as $k$ increases, $\alpha$ decreases. This is expected because a higher $k$ causes the ejecta to decelerate slower (assuming that $A$ in the equation $\rho=AR^{-k}$, which represents the density normalization at a characteristic radius, remains constant), resulting in a slower decline in $\theta_{\text{min}}(t)$ over time, leading to slower rising de-beamed emission.

After shock crossing, the temporal slope $\alpha$ becomes smaller than the pre-shock-crossing value by $2/a$ (an analytical expression is provided in Table~\ref{tab:temporal_indices}). This reduction is expected because the emission from each $\theta$ begins to decline after shock crossing, leading to de-beamed emission with smaller temporal slope. Aside from this decrease, the qualitative behavior of $\alpha$ after shock crossing is similar to its behavior before shock crossing, described by Eq.~(\ref{eq:alpha_FS_b}). Similar to the pre-shock crossing phase, $\alpha$ decreases as $k$ increases, as evident from the comparison between the $k=0$ case (solid red line) and the $k=2$ case (solid magenta line) in Fig.~\ref{fig:FS}.

The reason why $\alpha$ for FS emission is independent of $b$, the parameter governing the angular profile of the initial Lorentz factor, is that the post-shock-crossing Lorentz factor (cf. Eq.~\ref{eq:gamma_vs_t}) does not depend on $b$ and is identical in both the thick and thin shell cases. As we will see below, this is not the case for RSs.

\subsection{Reverse shock}
\label{sec:rs_emission}

\begin{figure}
\centering
\includegraphics[width=0.49\linewidth]{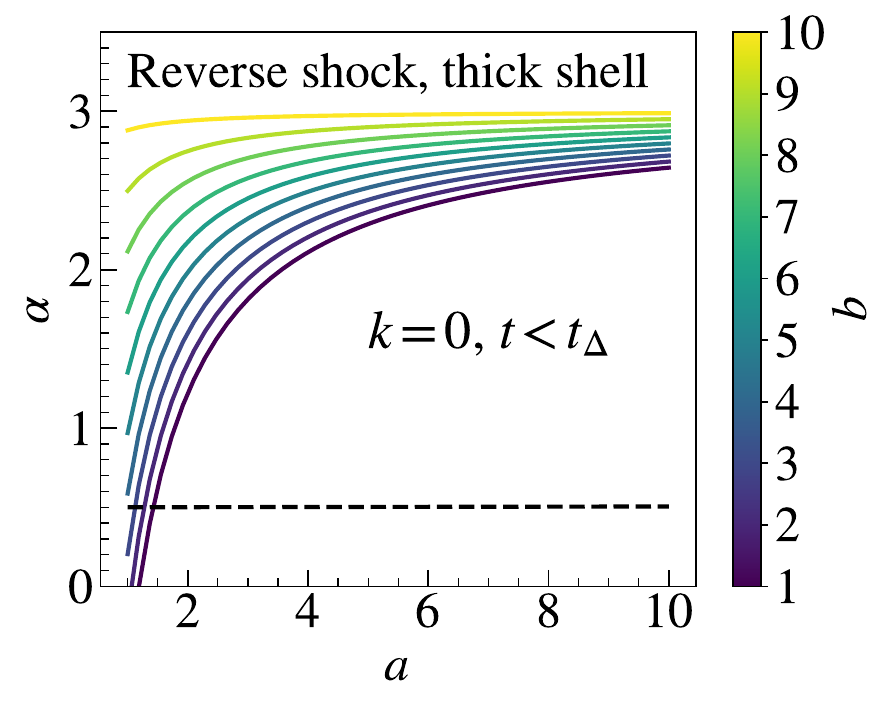}
\includegraphics[width=0.49\linewidth]{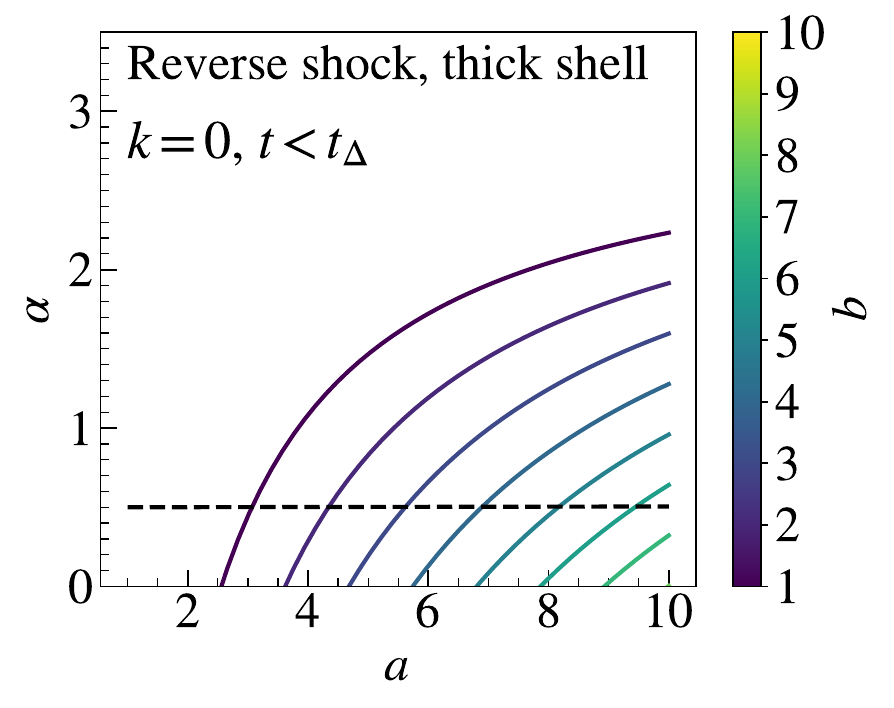}
\caption{Temporal slope $\alpha$ of de-beamed emission from the thick shell RS before shock crossing ($k = 0$) as a function of $a$. Curves represent different $b$ values, as indicated by the color bar. The horizontal dashed line marks the temporal slope for an on-axis observer and represents a minimum on the observable value of $\alpha$. The left panel corresponds to $\nu_\mathrm{m} < \nu < \nu_\mathrm{c}$, and the right panel to $\nu < \nu_\mathrm{m}$. While $\alpha$ increases with $a$ in both regimes, it grows (decreases) with $b$ in the left (right) panel.} 
\label{fig:RS_b}
\end{figure}

\begin{figure}
\centering
\includegraphics[width=0.49\linewidth]{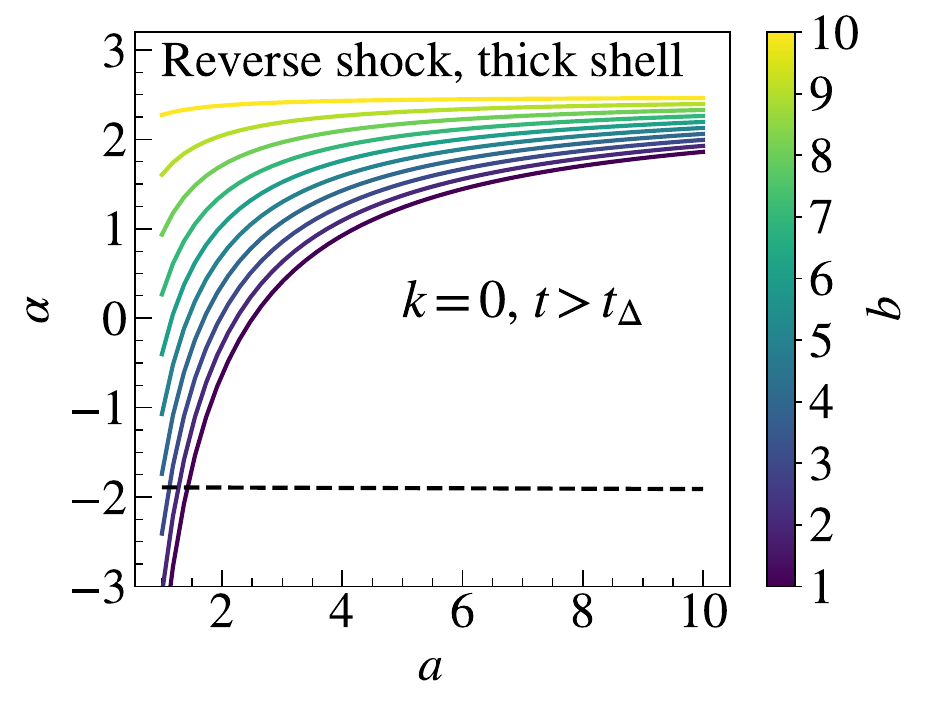}
\includegraphics[width=0.49\linewidth]{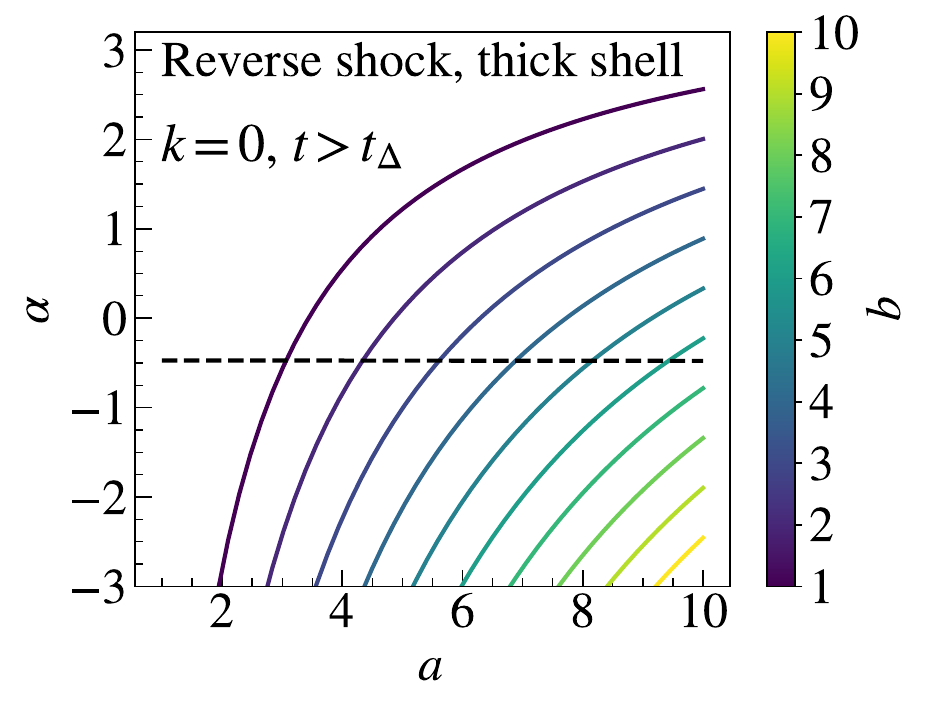}
\includegraphics[width=0.49\linewidth]{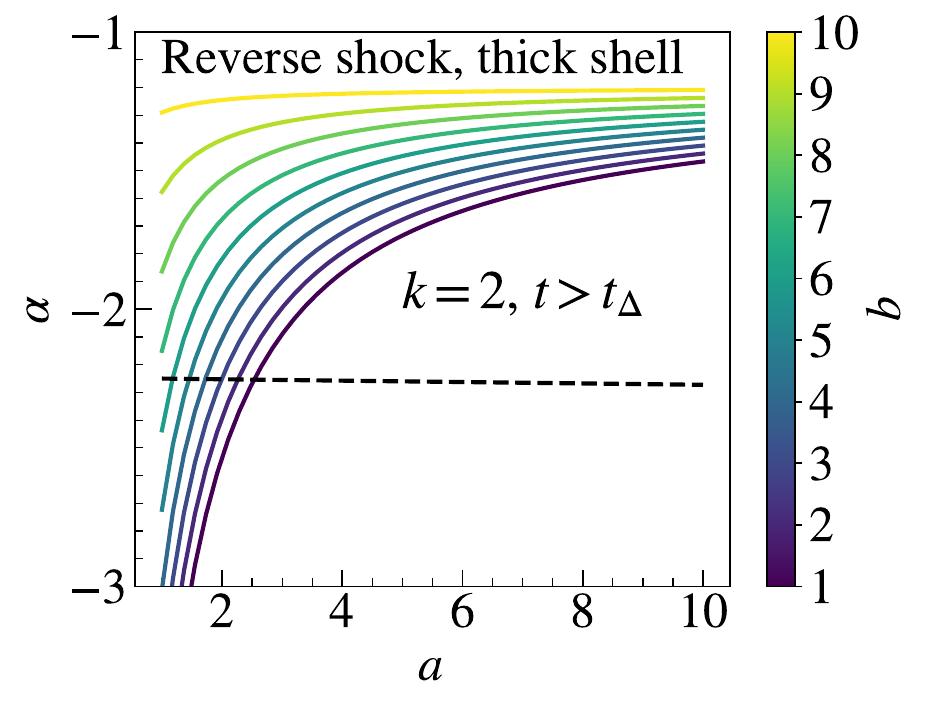}
\includegraphics[width=0.49\linewidth]{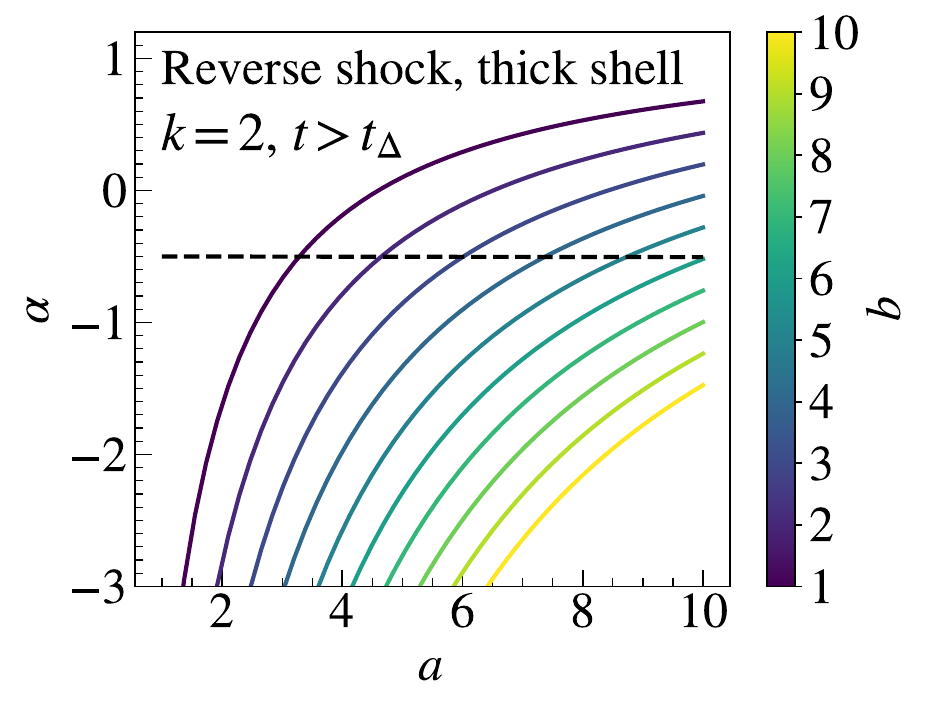}
\caption{Temporal slope $\alpha$ of the de-beamed emission from the thick shell RS after the shock crossing as a function of $a$. Curves represent different $b$ values, as indicated by the color bar. The horizontal dashed line marks the temporal slope for an on-axis observer and represents a minimum on the observable value of $\alpha$. The left panel corresponds to $\nu_\mathrm{m} < \nu < \nu_\mathrm{c}$, and the right panel to $\nu < \nu_\mathrm{m}$. The top and bottom panels correspond to $k = 0$ and $k = 2$, respectively. In all cases, $\alpha$ increases with $a$, but the dependence on $b$ is more complicated: it increases (decreases) with $b$ in the left (right) panels.} 
\label{fig:RS_thick_a}
\end{figure}

\begin{figure}
\centering
\includegraphics[width=0.49\linewidth]{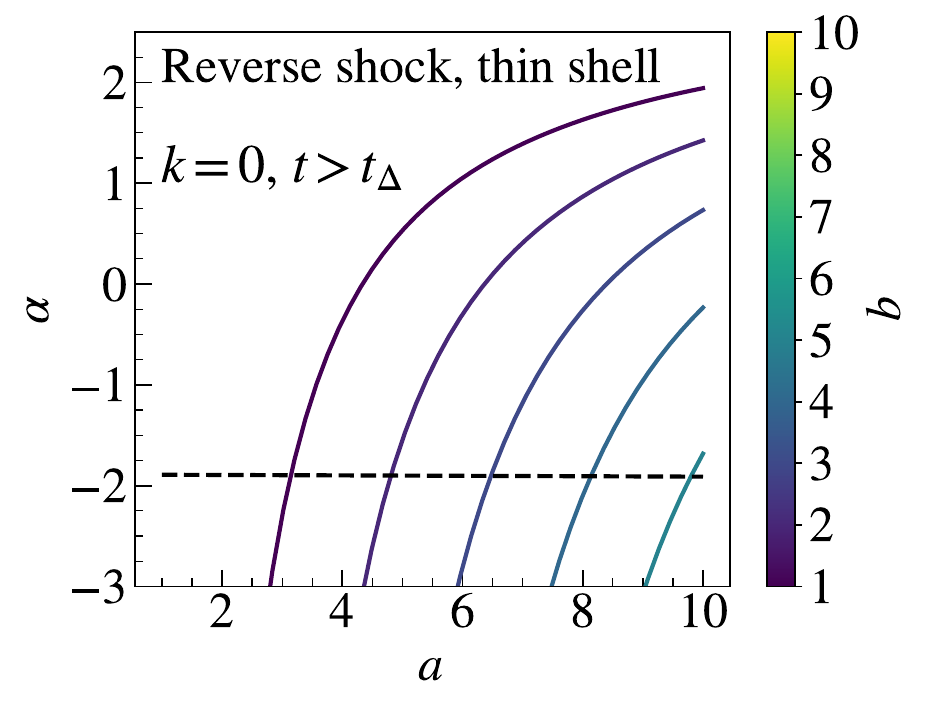}
\includegraphics[width=0.49\linewidth]{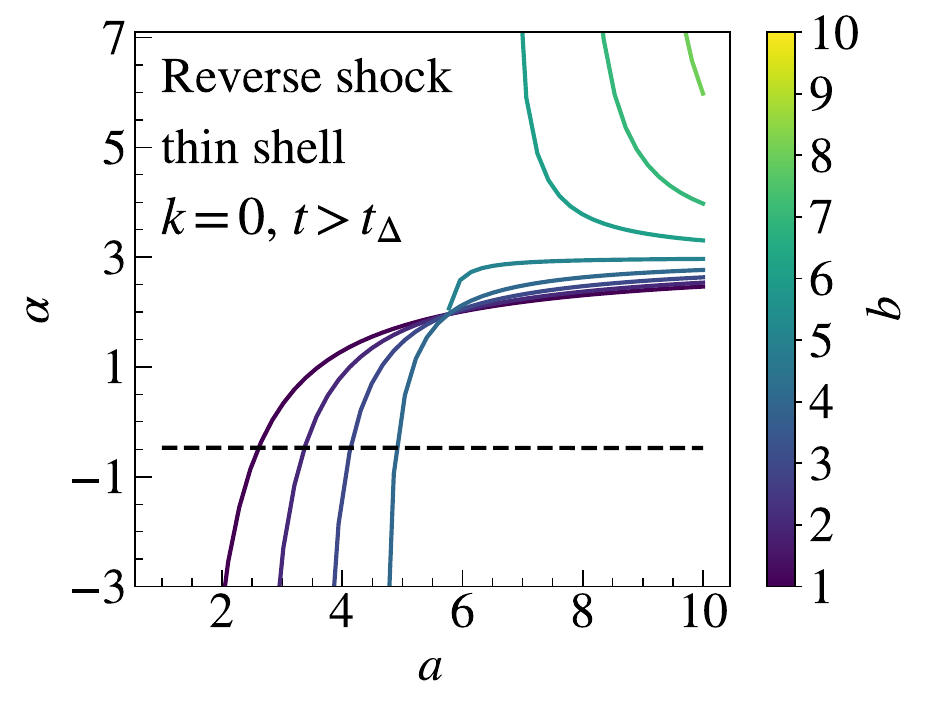}
\includegraphics[width=0.49\linewidth]{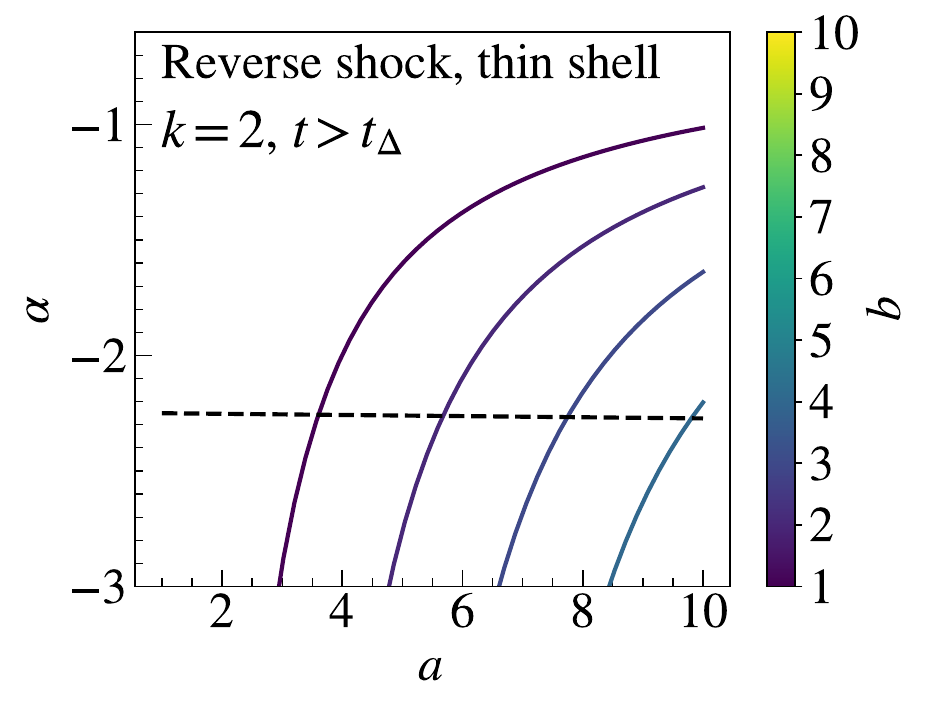}
\includegraphics[width=0.49\linewidth]{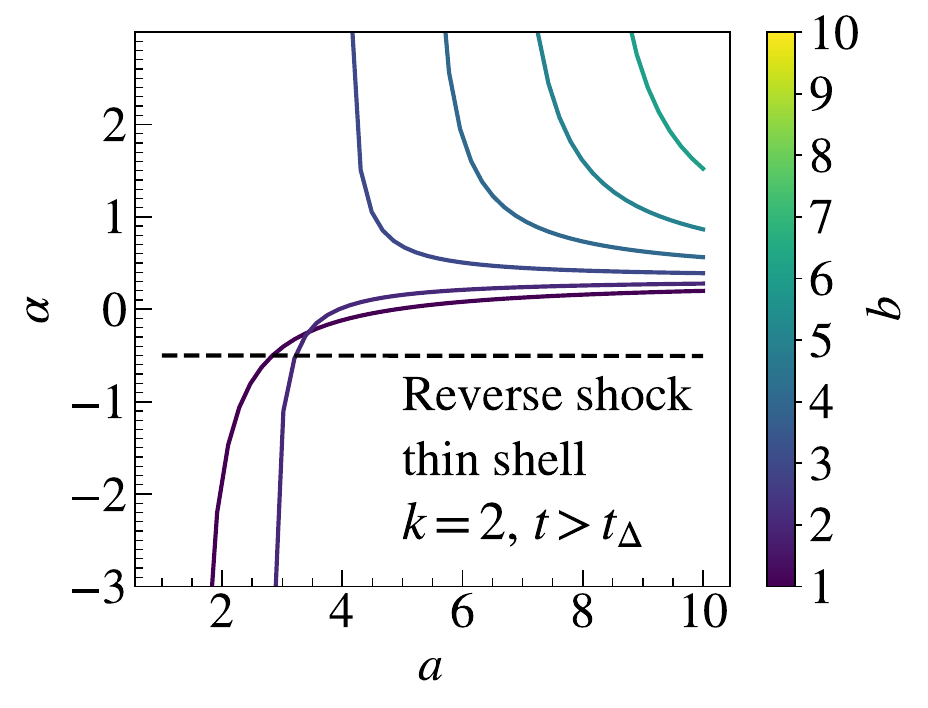}
\caption{Temporal slope $\alpha$ of the de-beamed emission from the thin shell RS after the shock crossing as a function of $a$. Curves represent different $b$ values, as indicated by the color bar. The horizontal dashed line marks the temporal slope for an on-axis observer and represents a minimum on the observable value of $\alpha$. The left panel corresponds to $\nu_\mathrm{m} < \nu < \nu_\mathrm{c}$, while the right panel to $\nu < \nu_\mathrm{m}$. The top and bottom panels correspond to $k = 0$ and $k = 2$, respectively. For spectral regime $\nu_\mathrm{m} < \nu < \nu_\mathrm{c}$, $\alpha$ increases with $a$, but for spectral regime $ \nu < \nu_\mathrm{m}$, the relationship is more complex.} 
\label{fig:RS_thin_a}
\end{figure}

Unlike FSs, the RS Lorentz factor evolution differs between thin and thick shells. For thin shells, the post-shock-crossing Lorentz factor depends on $b$ (Table~\ref{tab:g_bp_values}), while in thick shells, it does not. However, shocked shell properties, such as energy density, still depend on the Lorentz factor relative to the initial value, which means the emission properties will also depend on $b$. Moreover, this introduces variations across spectral regimes. This makes RS emission far more complex than FS emission \citep[e.g.,][]{Zhang24BOAT}.

Before shock crossing, the temporal slope $\alpha$ for thick shell RS is
\begin{equation}
    \alpha = -\frac{(k-2)(b(p-2)-2)}{a}-\frac{1}{4}k(p+5)+3,
\end{equation}
which is shown in Fig.~\ref{fig:RS_b} as a function of $a$ for $k=0$. Each solid curve corresponds to a different value of $b$, indicated by the color bar. For comparison, the dashed line represents the temporal slope for an on-axis observer. Similar to FS emission, $\alpha$ increases with $a$ for all $b$, as steeper jets have a higher angular gradient of luminosity, yielding higher temporal slopes. Unlike FS emission, $\alpha$ for the RS strongly depends on $b$. Moreover, the $b$ dependence varies by spectral regime: for $\nu_\mathrm{m} < \nu < \nu_\mathrm{c}$ (left panel), $\alpha$ grows with $b$ and exceeds the on-axis value for most values of $a$ and $b$. For $\nu < \nu_\mathrm{m}$ (right panel), $\alpha$ decreases with $b$. As a result, $\alpha$ becomes smaller than the corresponding on-axis value for large values of $b$. The contrasting dependence on $b$ in the two spectral regimes arises from the dependence on frequency $\nu_\mathrm{m}$: $F_{\nu}(\nu<\nu_{\rm m})\propto \nu_\mathrm{m}^{-1/3} \propto \Gamma_0^{-2/3}$ vs. $F_{\nu}(\nu_{\rm m}<\nu<\nu_{\rm c})\propto \nu_\mathrm{m}^{(p-1)/2} \propto \Gamma_0^{p-1} $. 

The temporal slope $\alpha$ for thick shell RS after shock crossing is shown in Fig.~\ref{fig:RS_thick_a}, an analytical expression is given in Table~\ref{tab:temporal_indices}. The overall dependence of $\alpha$ on $a$ and $b$ is qualitatively similar to that of the thick shell RS prior to shock crossing, discussed earlier. Similarly, $\alpha$ increases with $b$ for $\nu_\mathrm{m} < \nu < \nu_\mathrm{c}$ (left panels) and decreases with $b$ for $\nu < \nu_\mathrm{m}$ (right panels). A comparison between the top and bottom panels, showing the $k=0$ and $k=2$ cases respectively, reveals that $\alpha$, similar to the FSs discussed above, is larger for the $k=0$ case than for the $k=2$ case.

For thin shell RS after shock crossing, $\alpha$ is shown in Fig.~\ref{fig:RS_thin_a}, while an analytical expression is given in Table~\ref{tab:temporal_indices}. In the $\nu_\mathrm{m} < \nu < \nu_\mathrm{c}$ regime (left panel), $\alpha$ increases with $a$ for all $b$ values but decreases with $b$ for a given $a$. For $\nu < \nu_\mathrm{m}$ (right panel), $\alpha$ increases with $a$ when $b < 6$, but decreases with $a$ for $b \geq 7$. Similar to the thick shell case, $\alpha$ is larger for $k=0$ (top panels) than for $k=2$ (bottom panels).

\begin{figure}
\centering
\includegraphics[width=0.49\linewidth]{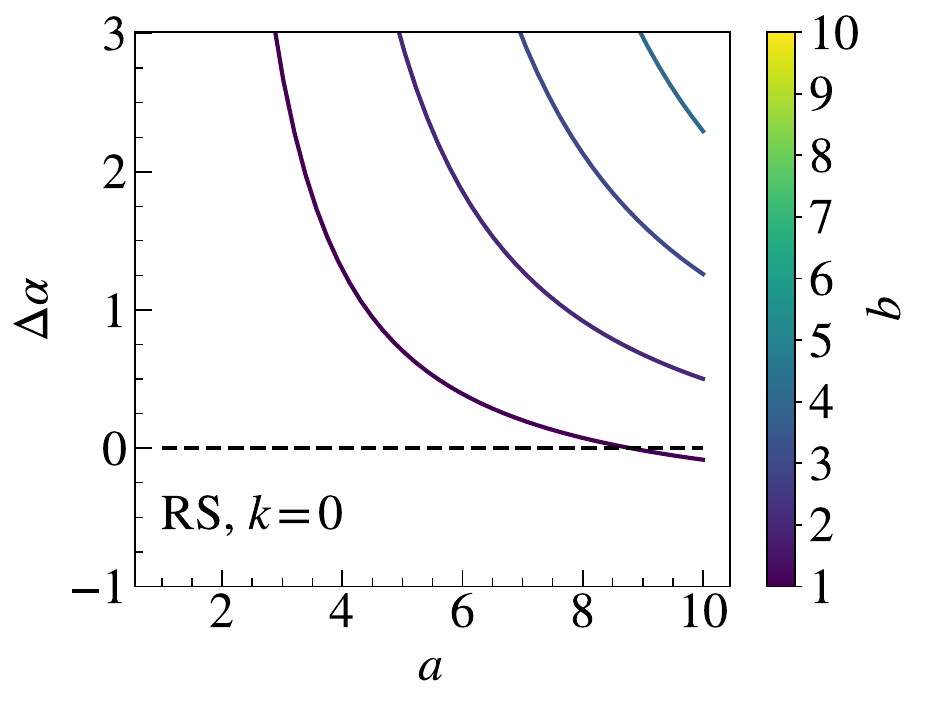}
\includegraphics[width=0.49\linewidth]{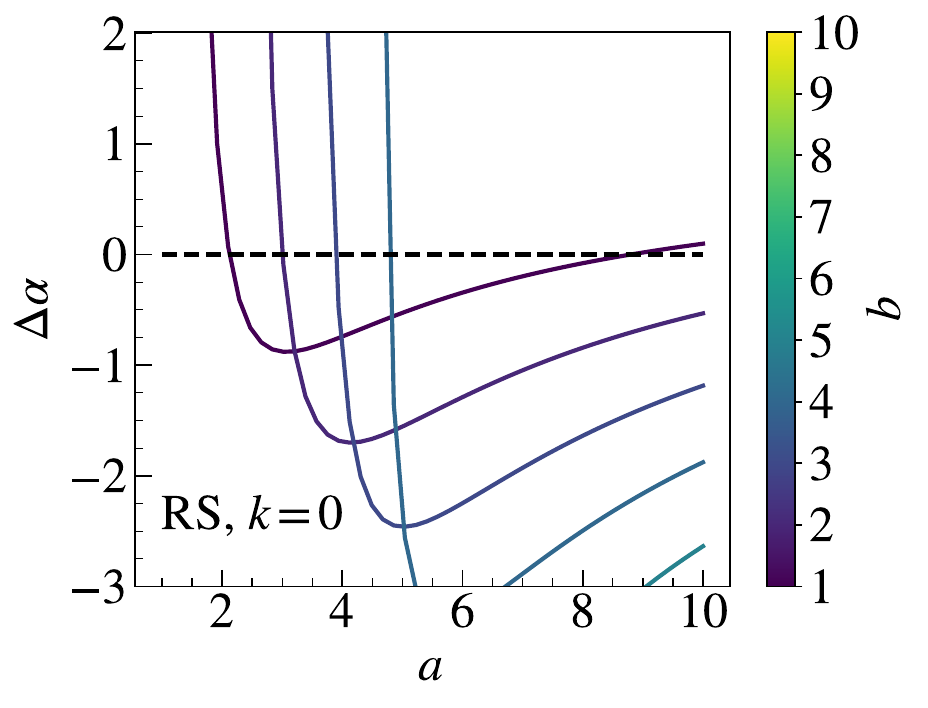}
\includegraphics[width=0.49\linewidth]{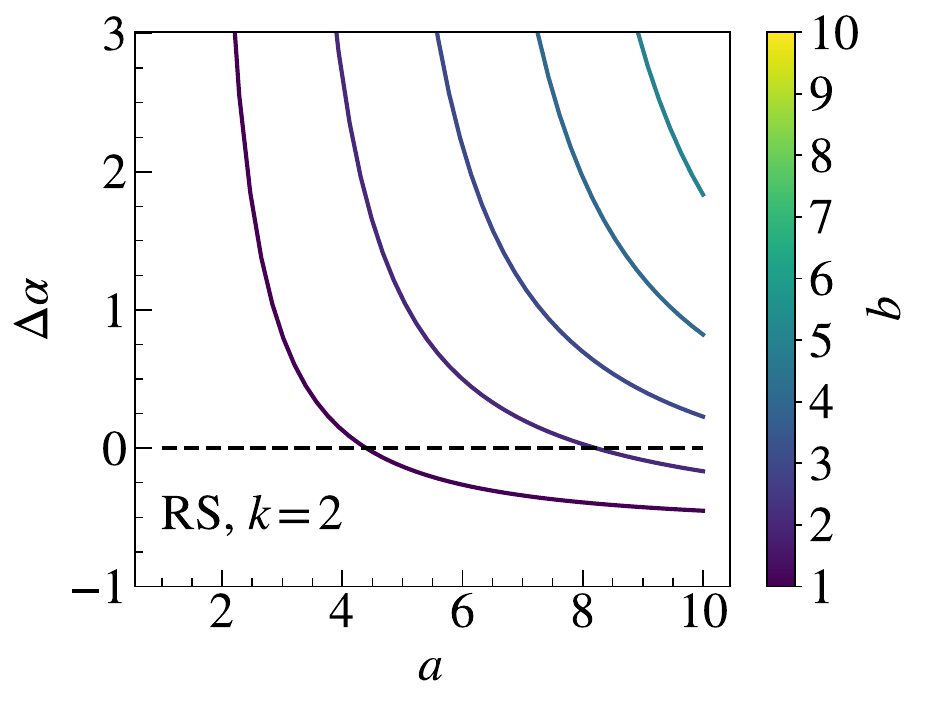}
\includegraphics[width=0.49\linewidth]{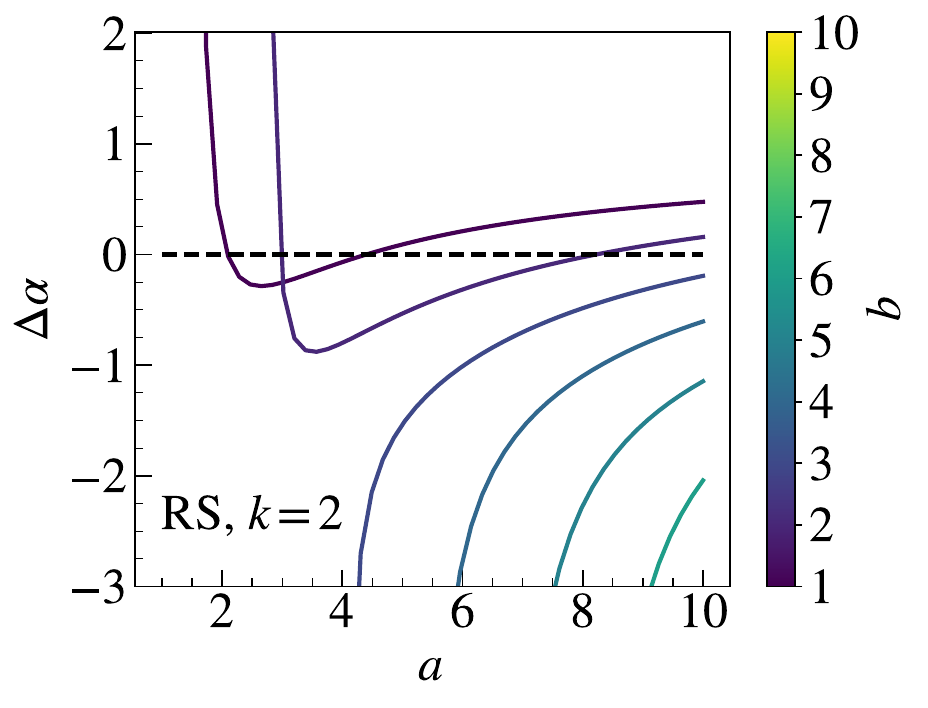}
\caption{Change of temporal slope $\alpha$ at the transition from thin to thick shell for RSs as a function of $a$. Curves represent different $b$ values, as indicated by the color bar. The left panel corresponds to $\nu_\mathrm{m} < \nu < \nu_\mathrm{c}$, and the right panel to $\nu < \nu_\mathrm{m}$. The top and bottom panels correspond to $k = 0$ and $k = 2$, respectively.} 
\label{fig:RS_RvsN_a}
\end{figure}

The change in the temporal slope, $\Delta \alpha$, at the transition from thin to thick shell is presented in Fig.~\ref{fig:RS_RvsN_a} as a function of $a$. For $\nu_\mathrm{m} < \nu < \nu_\mathrm{c}$ (left panel), $\Delta \alpha$ is positive for most values of $a$ and $b$, indicating that the temporal slope increases at the transition. The corresponding light curve behavior for cases B and C is illustrated by the solid blue and purple lines in the bottom center and right panels of Fig.~\ref{fig:LC_cases}, respectively. In contrast, for $\nu < \nu_\mathrm{m}$ (right panel), $\Delta \alpha$ can become negative, particularly for large $a$ and $b$, signifying a decreasing temporal slope at the transition. The corresponding light curve behavior is illustrated with dashed blue and purple lines in the bottom center and right panels of Fig.~\ref{fig:LC_cases} for cases B and C, respectively.

\begin{figure}
\centering
\includegraphics[width=0.49\linewidth]{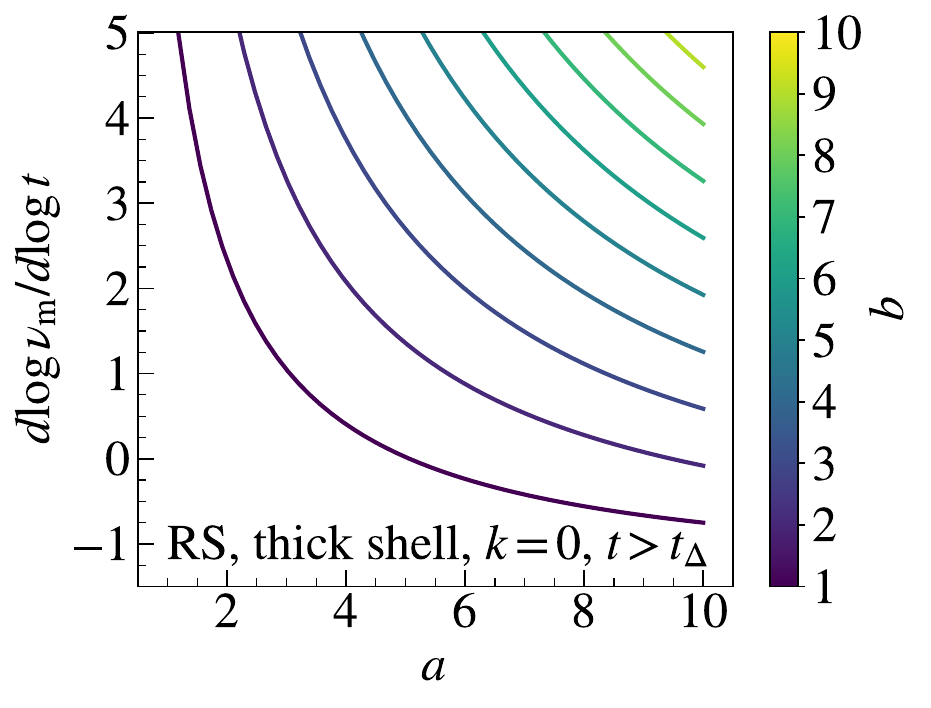}
\includegraphics[width=0.49\linewidth]{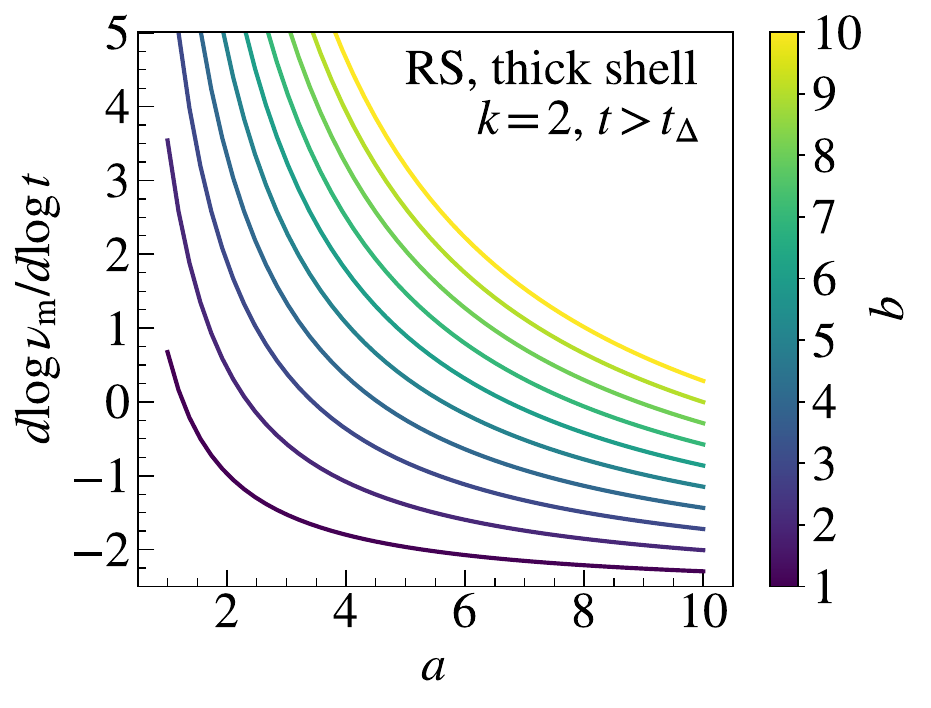}
\includegraphics[width=0.49\linewidth]{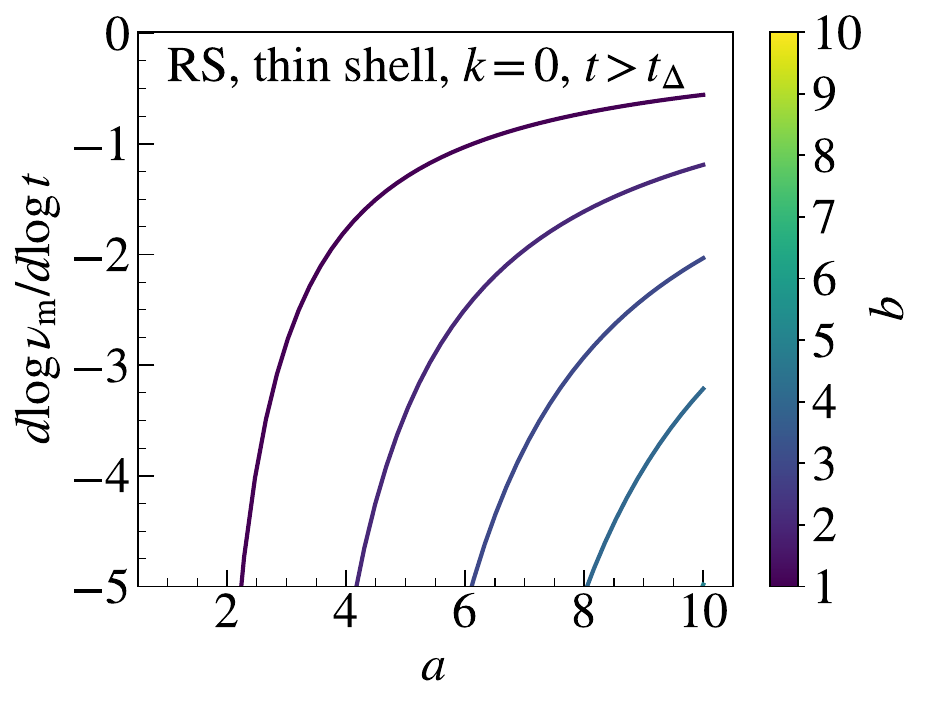}
\includegraphics[width=0.49\linewidth]{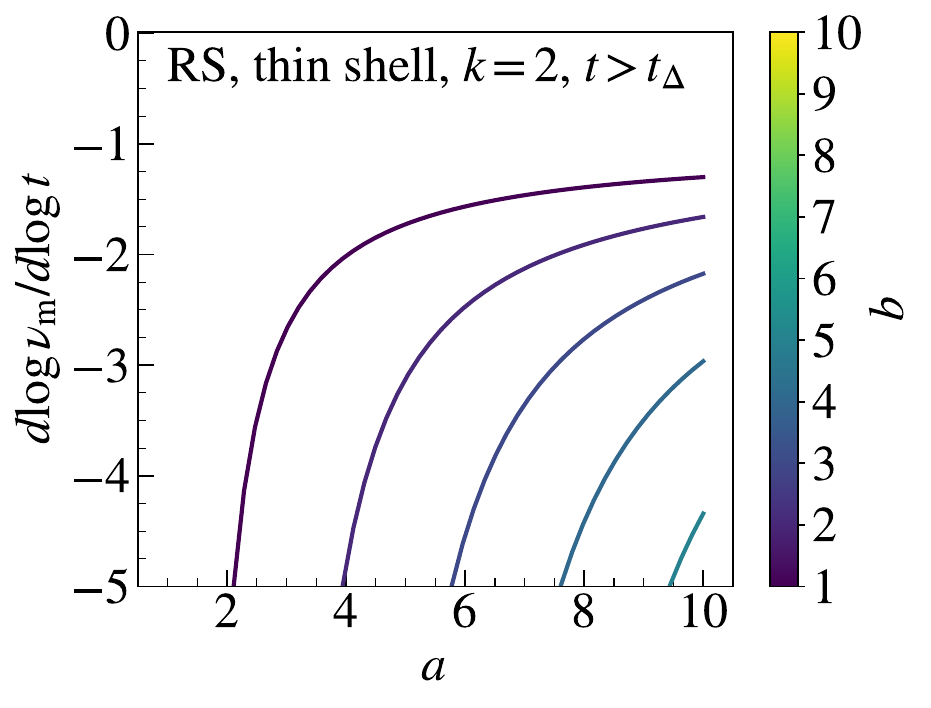}
\caption{The temporal slope of $\nu_\mathrm{m}(t)$ during de-beamed emission after shock crossing for RS as a function $a$. Curves represent different $b$ values, as indicated by the color bar. The left panel corresponds to $k=0$, while the right panel shows $k=2$ case. The top and bottom panels correspond to thick and thin shells, respectively.} 
\label{fig:num_RS}
\end{figure}

It is insightful to examine the evolution of the frequency $\nu_\mathrm{m}(t)$ during the de-beamed emission. For FS, it evolves as $t^{-k/2}$, before and after shock crossing for both thin and thick shells. The situation is a lot of more complex for RSs. Fig.~\ref{fig:num_RS} shows the temporal slope of $\nu_\mathrm{m}(t)$ for RSs as a function $a$ for different $b$ after shock crossing. The top and bottom panels show thick and thin shells, respectively. The magnitude of the slope decrease with increasing $a$. The same trend happens when $b$ decreases. As a result, the spectral evolution of de-beamed emission from RSs is slow for small $b$ and large $a$. This could be consistent with the slow spectral evolution observed in, e.g., GRB 130427A \citep{Laskar13Reverse, Perley14Afterglow}, 181201A \citep{Laskar19Reverse}, and 221009A \citep{Laskar23Radio, Bright23Precise}. However, drawing definitive conclusions from our simplified analytical model would be premature. A more detailed numerical study is necessary to rigorously verify these results.  

\subsection{Light curve examples}
\label{sec:lc_example}

In this section, we present light curve examples for cases A, B, and C for specific jet parameters. To construct light curves, we employ piecewise power-law segments with smooth transitions between them, following \cite{beniamini20afterglow}. We focus on the spectral regime $\nu_\mathrm{m} < \nu < \nu_\mathrm{c}$. Although afterglow emission from real GRBs likely transitions between different spectral regimes \citep[e.g.,][]{Wang15How}, for simplicity, we neglect these transitions here. The peak of the light curve, corresponding to the moment when the jet core becomes visible, is calculated using the approach of \cite{Nakar02Detectability}, which for $p=2.2$ is given by
\begin{align}
F_\mathrm{pk} \approx 620 \epsilon_{e,-1}^{p-1} \epsilon_{B,-2}^{(p+1)/4} n_0^{(p+1)/4}
E_{50.7} \nu_{14.7}^{(1-p)/2} \\\nonumber
\times D_{L28}^{-2} \theta_{\mathrm{obs},-1}^{-2p} (1+z)^{(3-p)/2} \, \mu\mathrm{Jy},
\end{align}
where $\epsilon_e$ and $\epsilon_B$ denote the fractions of shock energy transferred to electrons and magnetic fields, respectively. Here, $Q_x$ represents the value of the quantity $Q$ expressed in $10^x$ cgs units. $D_L$ is the luminosity distance, $n$ is the ISM density, $E$ is the jet energy, and $\nu$ is the observed frequency. We focus on jet with $a=4$, $b=2$, and $\theta_\mathrm{c}=0.03$. The ratio of  $\epsilon_B$ for FS to that of RS is assumed to be $4$, which is consistent with observations \citep[e.g.,][]{Gao15Morphological}. Although this analytical approach clearly reveals the underlying physical dependencies, it lacks the precision of numerical approaches in predicting light curves for given parameter sets. Moreover, the range of possible light curve morphologies is extensive, and we examine only a single realization for each of cases A, B, and C. Thus, the results below should be considered with caution.

\begin{figure*}
\centering
\includegraphics[width=0.32\linewidth]{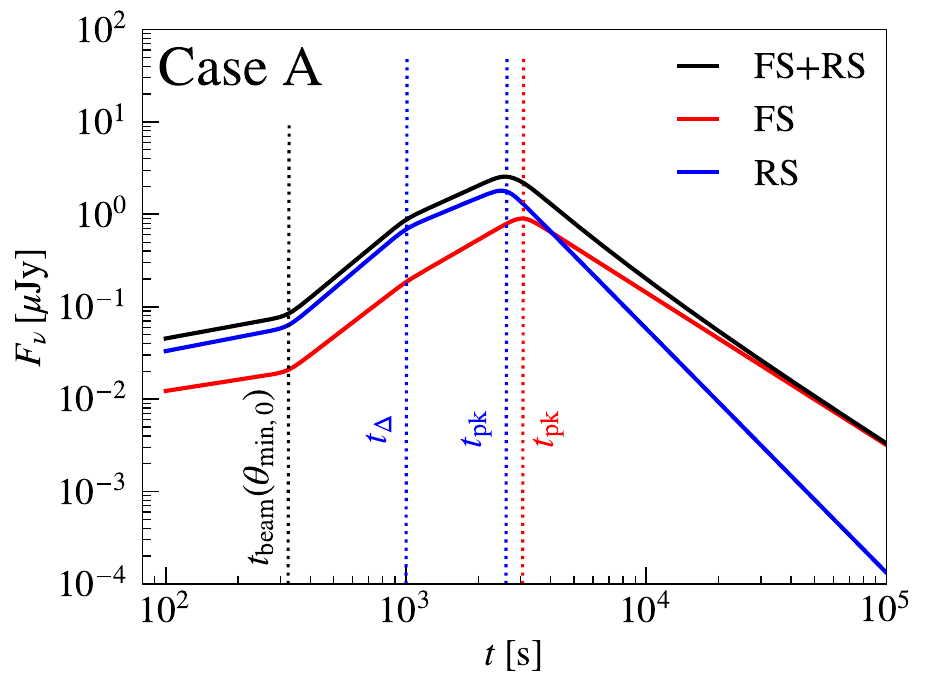}
\includegraphics[width=0.32\linewidth]{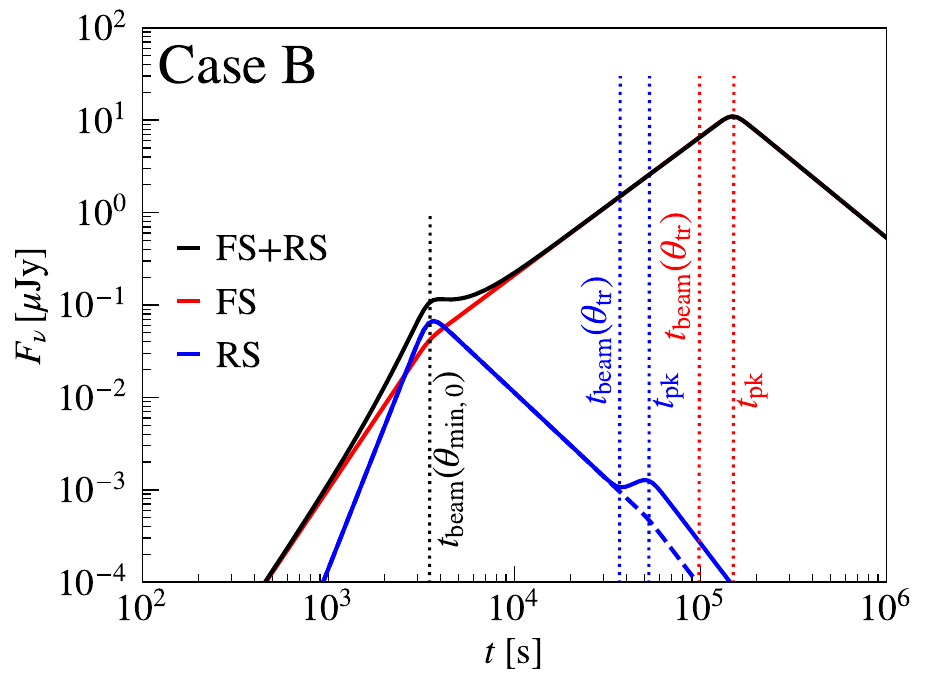}
\includegraphics[width=0.32\linewidth]{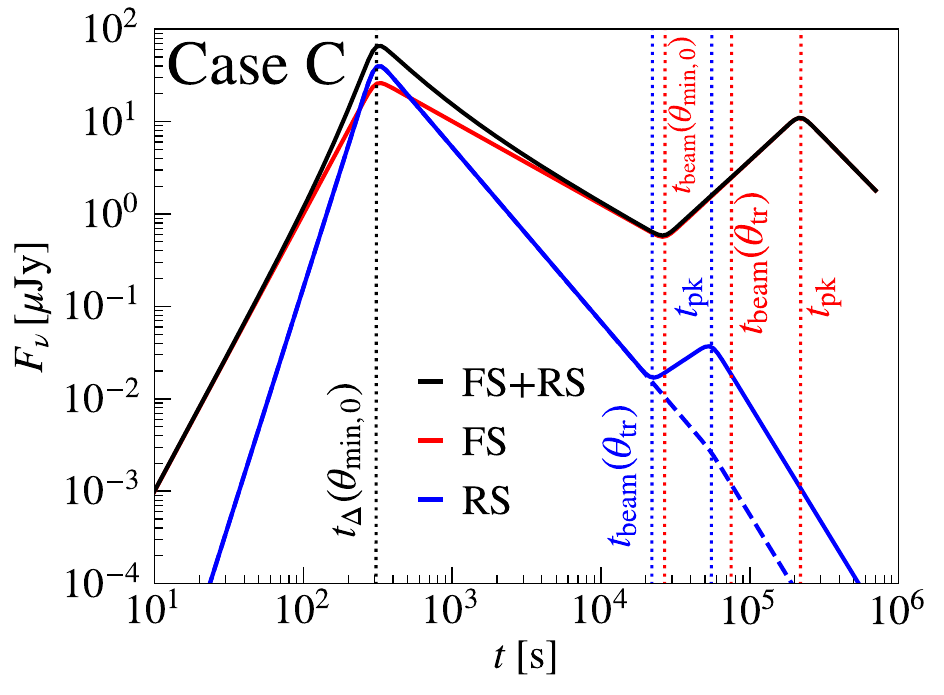}
\caption{Light curves for cases A, B, and C at frequency $\nu=10^{14.7}$ Hz. Blue (red) legends denote FS (RS), while black legends apply to both. Jet parameters are $a=4$, $b=2$, $\theta_\mathrm{c}=0.03$. Case A parameters: energy $E=10^{48} \, \mathrm{erg}$, shell width $\Delta_0=3\times10^{13} \, \mathrm{cm}$, $\Gamma_\mathrm{c,0}=200$, and $\theta_\mathrm{obs}=0.17$. Case B: $E=10^{50.7} \, \mathrm{erg}$, $\Delta_0=3\times10^{12} \, \mathrm{cm}$, $\Gamma_\mathrm{c,0}=200$, and $\theta_\mathrm{tr}=0.05$. Case C: $E=10^{50.7} \, \mathrm{erg}$, $\Delta_0=4.2\times10^{11} \,\mathrm{cm}$, $\Gamma_\mathrm{c,0}=1000$, and $\theta_\mathrm{tr}=0.05$. For B and C, $\theta_\mathrm{obs}=0.25$. Dashed blue lines represent cases without transition to a thick shell in the jet core.}
\label{fig:LC_example}
\end{figure*}

The left panel of Fig.~\ref{fig:LC_example} shows the light curve at $\nu=10^{14.7}$ Hz for case A. As mentioned above, this case is expected in bursts with low energy, long duration, and dense circumstellar environments. This model assumes $E=10^{48} \,\mathrm{erg}$, $\Delta_0=3\times10^{13} \,\mathrm{cm}$, $\Gamma_\mathrm{c,0}=200$, and $\theta_\mathrm{obs}=0.17$. Due to significant deceleration before shock crossing, the de-beamed emission commences around $300$ s, well before the RS traverses the shell at $10^3$ s. This de-beamed emission persists until the jet core becomes visible at $\sim 3000$ s. Up to this point, RS emission dominates over FS emission. Beyond this, FS emission becomes dominant.

The center panel of Fig.~\ref{fig:LC_example} shows the light curve for case B. This corresponds to a jet with $E=10^{50.7} \,\mathrm{erg}$, $\Delta_0=3\times10^{12} \,\mathrm{cm}$, $\Gamma_\mathrm{c,0}=200$, $\theta_\mathrm{obs}=0.25$, and transition angle $\theta_\mathrm{tr}=0.05$. The RS crosses the shell at $\theta_\mathrm{min,0}$ around $3.5\times 10^3$ s, triggering de-beamed emission. The FS emission rises, while the RS emission declines (RS can exhibit rising de-beamed emission for steeper jets, e.g., for $a=7$, $b=2$). After shock crossing, FS emission dominates over RS. Around $4\times 10^4$ s, RS emission transitions to the thick shell core but remains buried under FS due to its weaker emission (the dashed blue lines show RS emission from purely thin shell jet). The FS transitions to thick shell core slightly later, but since both thin- and thick-shell phases exhibit similar temporal slopes for de-beamed emission, this transition is difficult to identity. Case C, shown in the right panel of Fig.~\ref{fig:LC_example}, has $E=10^{50.7} \,\mathrm{erg}$, $\Delta_0=4.2\times10^{11} \,\mathrm{cm}$, $\Gamma_\mathrm{c,0}=1000$, $\theta_\mathrm{obs}=0.25$, and $\theta_\mathrm{tr}=0.05$. The light curve from FS features double peaks, as de-beamed emission starts well after shock crossing \citep{beniamini20afterglow}. As in case B, the RS emission remains buried under FS emission after shock crossing.

\section{Conclusions}
\label{sec:conclusion}

In this study, we explored the morphology of gamma-ray burst (GRB) afterglows produced by steep jets observed off-axis. Using an analytical framework, we conducted an extensive parameter study. As the jet decelerates, relativistic beaming weakens, revealing the inner, more luminous regions. The minimum visible angle decreases either before, at, or after the reverse shock (RS) crosses the ejecta, leading to three distinct light curve morphologies. We analyzed these morphologies for both forward shocks (FS) and RS.

We demonstrated that de-beamed emission can produce rapidly rising signals even before the RS crosses the ejecta, in jets that decelerate significantly before shock crossing. This is expected in GRBs with very long duration, low energy, or a dense circum-burst medium (or combinations thereof). For FSs, the temporal slopes of de-beamed emission are independent of the initial angular Lorentz factor profile and shell type (thick or thin). In contrast, RS emission depends on both, resulting in diverse light curve variations. 

For FS, the characteristic synchrotron frequency $\nu_\mathrm{m}(t)$ evolves as $t^{-k/2}$ during de-beamed emission. It is independent of the jet structure. For RS, it depends on the lateral steepness of both energy and Lorentz factor. A steep energy and shallow Lorentz factor gradients yield slow evolution.

Ejecta can behave as a thick shell in the inner regions of the jet and a thin shell in the outer regions. For FSs, since the temporal slopes remain identical across these regimes, transitions are hard to identify in the observed signal. For RSs, thick and thin shell emissions often exhibit distinct temporal slopes. If thick shell RS emission exceeds that of thin shell RS, the transition may manifest as a hump in the light curve if the RS emission manages to rise above the FS emission. 

In addition to the limitations outlined in Section~\ref{sec:method}, our work is subject to several others. While analytical methods enable detailed parameter studies and effectively highlight the underlying physical principles, they lack the precision required to accurately compute light curves for specific parameter values. Therefore, it is essential to compare our results with detailed numerical simulations. Additionally, validation against multi-wavelength observations, particularly during the early phase of the afterglow \citep[e.g.,][]{Vestrand14Bright, Xin23Prompt, Becerra23Understanding, Komesh23Evolution, Sadeh24Detecting}, is crucial to verify our predictions for de-beamed emission before shock crossing. Furthermore, our analysis is restricted to steep jets, whereas some GRBs may exhibit shallow angular structures \citep{Beniamini22Robust, Gill23GRB}. We disregard jet spreading, which can be important in later stages of the evolution \citep[e.g.,][]{vanEerten12Gamma}. We neglect self-absorption, which may be significant, particularly for radio signals \citep[e.g.,][]{Bright23Precise}. We also focus on matter-dominated jets, whereas some jets may be Poynting flux-dominated \citep[e.g.,][]{Zhang18Transition} or possess a mixed structure, e.g. with a Poynting flux-dominated core and a matter-dominated wing \citep{Zhang24BOAT}. These limitations will be addressed in future works.

\section*{Acknowledgements}

We thank Tanmoy Laskar for useful discussion. This work is partially supported by Nazarbayev University Faculty Development Competitive Research Grant Program (no. 040225FD4713). PB's work was funded by a NASA grant 80NSSC24K0770, a grant (no. 2020747) from the United States-Israel Binational Science Foundation (BSF), Jerusalem, Israel and by a grant (no. 1649/23) from the Israel Science Foundation.

\section*{Data Availability}

The data used in this work is available from authors upon request. 
 

    \bibliographystyle{mnras}

\begin{thebibliography}{}
    \makeatletter
    \relax
    \def\mn@urlcharsother{\let\do\@makeother \do\$\do\&\do\#\do\^\do\_\do\%\do\~}
    \def\mn@doi{\begingroup\mn@urlcharsother \@ifnextchar [ {\mn@doi@} {\mn@doi@[]}}
    \def\mn@doi@[#1]#2{\def\@tempa{#1}\ifx\@tempa\@empty \href {http://dx.doi.org/#2} {doi:#2}\else \href {http://dx.doi.org/#2} {#1}\fi \endgroup}
    \def\mn@eprint#1#2{\mn@eprint@#1:#2::\@nil}
    \def\mn@eprint@arXiv#1{\href {http://arxiv.org/abs/#1} {{\tt arXiv:#1}}}
    \def\mn@eprint@dblp#1{\href {http://dblp.uni-trier.de/rec/bibtex/#1.xml} {dblp:#1}}
    \def\mn@eprint@#1:#2:#3:#4\@nil{\def\@tempa {#1}\def\@tempb {#2}\def\@tempc {#3}\ifx \@tempc \@empty \let \@tempc \@tempb \let \@tempb \@tempa \fi \ifx \@tempb \@empty \def\@tempb {arXiv}\fi \@ifundefined {mn@eprint@\@tempb}{\@tempb:\@tempc}{\expandafter \expandafter \csname mn@eprint@\@tempb\endcsname \expandafter{\@tempc}}}
    
    \bibitem[\protect\citeauthoryear{{Abbott} et~al.,}{{Abbott} et~al.}{2017}]{abbott2017gw170817b}
    {Abbott} B.~P.,  et~al., 2017, \mn@doi [\apjl] {10.3847/2041-8213/aa91c9}, \href {https://ui.adsabs.harvard.edu/abs/2017ApJ...848L..12A} {848, L12}
    
    \bibitem[\protect\citeauthoryear{{Alexander} et~al.,}{{Alexander} et~al.}{2018}]{Alexander18Decline}
    {Alexander} K.~D.,  et~al., 2018, \mn@doi [\apjl] {10.3847/2041-8213/aad637}, \href {https://ui.adsabs.harvard.edu/abs/2018ApJ...863L..18A} {863, L18}
    
    \bibitem[\protect\citeauthoryear{{Becerra} et~al.,}{{Becerra} et~al.}{2023}]{Becerra23Understanding}
    {Becerra} R.~L.,  et~al., 2023, \mn@doi [\mnras] {10.1093/mnras/stad2513}, \href {https://ui.adsabs.harvard.edu/abs/2023MNRAS.525.3262B} {525, 3262}
    
    \bibitem[\protect\citeauthoryear{{Beniamini} \& {Nakar}}{{Beniamini} \& {Nakar}}{2019}]{Beniamini19Observational}
    {Beniamini} P.,  {Nakar} E.,  2019, \mn@doi [\mnras] {10.1093/mnras/sty3110}, \href {https://ui.adsabs.harvard.edu/abs/2019MNRAS.482.5430B} {482, 5430}
    
    \bibitem[\protect\citeauthoryear{{Beniamini}, {Petropoulou}, {Barniol Duran}  \& {Giannios}}{{Beniamini} et~al.}{2019}]{Beniamini2019}
    {Beniamini} P.,  {Petropoulou} M.,  {Barniol Duran} R.,   {Giannios} D.,  2019, \mn@doi [\mnras] {10.1093/mnras/sty3093}, \href {https://ui.adsabs.harvard.edu/abs/2019MNRAS.483..840B} {483, 840}
    
    \bibitem[\protect\citeauthoryear{{Beniamini}, {Granot}  \& {Gill}}{{Beniamini} et~al.}{2020}]{beniamini20afterglow}
    {Beniamini} P.,  {Granot} J.,   {Gill} R.,  2020, \mn@doi [\mnras] {10.1093/mnras/staa538}, \href {https://ui.adsabs.harvard.edu/abs/2020MNRAS.493.3521B} {493, 3521}
    
    \bibitem[\protect\citeauthoryear{{Beniamini}, {Gill}  \& {Granot}}{{Beniamini} et~al.}{2022}]{Beniamini22Robust}
    {Beniamini} P.,  {Gill} R.,   {Granot} J.,  2022, \mn@doi [\mnras] {10.1093/mnras/stac1821}, \href {https://ui.adsabs.harvard.edu/abs/2022MNRAS.515..555B} {515, 555}
    
    \bibitem[\protect\citeauthoryear{{Beniamini}, {Piran}  \& {Matsumoto}}{{Beniamini} et~al.}{2023}]{Beniamini23Swift}
    {Beniamini} P.,  {Piran} T.,   {Matsumoto} T.,  2023, \mn@doi [\mnras] {10.1093/mnras/stad1950}, \href {https://ui.adsabs.harvard.edu/abs/2023MNRAS.524.1386B} {524, 1386}
    
    \bibitem[\protect\citeauthoryear{{Blandford} \& {McKee}}{{Blandford} \& {McKee}}{1976}]{blandford76fluid}
    {Blandford} R.~D.,  {McKee} C.~F.,  1976, \mn@doi [Physics of Fluids] {10.1063/1.861619}, \href {https://ui.adsabs.harvard.edu/abs/1976PhFl...19.1130B} {19, 1130}
    
    \bibitem[\protect\citeauthoryear{{Bright} et~al.,}{{Bright} et~al.}{2023}]{Bright23Precise}
    {Bright} J.~S.,  et~al., 2023, \mn@doi [Nature Astronomy] {10.1038/s41550-023-01997-9}, \href {https://ui.adsabs.harvard.edu/abs/2023NatAs...7..986B} {7, 986}
    
    \bibitem[\protect\citeauthoryear{{De Colle}, {Granot}, {L{\'o}pez-C{\'a}mara}  \& {Ramirez-Ruiz}}{{De Colle} et~al.}{2012a}]{DeColle12Gamma_1}
    {De Colle} F.,  {Granot} J.,  {L{\'o}pez-C{\'a}mara} D.,   {Ramirez-Ruiz} E.,  2012a, \mn@doi [\apj] {10.1088/0004-637X/746/2/122}, \href {https://ui.adsabs.harvard.edu/abs/2012ApJ...746..122D} {746, 122}
    
    \bibitem[\protect\citeauthoryear{{De Colle}, {Ramirez-Ruiz}, {Granot}  \& {Lopez-Camara}}{{De Colle} et~al.}{2012b}]{DeColle12Simulations}
    {De Colle} F.,  {Ramirez-Ruiz} E.,  {Granot} J.,   {Lopez-Camara} D.,  2012b, \mn@doi [\apj] {10.1088/0004-637X/751/1/57}, \href {https://ui.adsabs.harvard.edu/abs/2012ApJ...751...57D} {751, 57}
    
    \bibitem[\protect\citeauthoryear{{Duque}, {Beniamini}, {Daigne}  \& {Mochkovitch}}{{Duque} et~al.}{2022}]{Duque22Flares}
    {Duque} R.,  {Beniamini} P.,  {Daigne} F.,   {Mochkovitch} R.,  2022, \mn@doi [\mnras] {10.1093/mnras/stac938}, \href {https://ui.adsabs.harvard.edu/abs/2022MNRAS.513..951D} {513, 951}
    
    \bibitem[\protect\citeauthoryear{{Eichler} \& {Granot}}{{Eichler} \& {Granot}}{2006}]{Eichler06Case}
    {Eichler} D.,  {Granot} J.,  2006, \mn@doi [\apjl] {10.1086/503667}, \href {https://ui.adsabs.harvard.edu/abs/2006ApJ...641L...5E} {641, L5}
    
    \bibitem[\protect\citeauthoryear{{Fraija}, {De Colle}, {Veres}, {Dichiara}, {Barniol Duran}, {Galvan-Gamez}  \& {Pedreira}}{{Fraija} et~al.}{2019}]{Fraija19Short}
    {Fraija} N.,  {De Colle} F.,  {Veres} P.,  {Dichiara} S.,  {Barniol Duran} R.,  {Galvan-Gamez} A.,   {Pedreira} A.~C. C. d. E.~S.,  2019, \mn@doi [\apj] {10.3847/1538-4357/aaf564}, \href {https://ui.adsabs.harvard.edu/abs/2019ApJ...871..123F} {871, 123}
    
    \bibitem[\protect\citeauthoryear{{Frail} et~al.,}{{Frail} et~al.}{2001}]{Frail01Beaming}
    {Frail} D.~A.,  et~al., 2001, \mn@doi [\apjl] {10.1086/338119}, \href {https://ui.adsabs.harvard.edu/abs/2001ApJ...562L..55F} {562, L55}
    
    \bibitem[\protect\citeauthoryear{{Gao} \& {M{\'e}sz{\'a}ros}}{{Gao} \& {M{\'e}sz{\'a}ros}}{2015}]{gao15reverse}
    {Gao} H.,  {M{\'e}sz{\'a}ros} P.,  2015, \mn@doi [Advances in Astronomy] {10.1155/2015/192383}, \href {https://ui.adsabs.harvard.edu/abs/2015AdAst2015E..13G} {2015, 192383}
    
    \bibitem[\protect\citeauthoryear{{Gao}, {Lei}, {Zou}, {Wu}  \& {Zhang}}{{Gao} et~al.}{2013}]{Gao13complete}
    {Gao} H.,  {Lei} W.-H.,  {Zou} Y.-C.,  {Wu} X.-F.,   {Zhang} B.,  2013, \mn@doi [\nar] {10.1016/j.newar.2013.10.001}, \href {https://ui.adsabs.harvard.edu/abs/2013NewAR..57..141G} {57, 141}
    
    \bibitem[\protect\citeauthoryear{{Gao}, {Wang}, {M{\'e}sz{\'a}ros}  \& {Zhang}}{{Gao} et~al.}{2015}]{Gao15Morphological}
    {Gao} H.,  {Wang} X.-G.,  {M{\'e}sz{\'a}ros} P.,   {Zhang} B.,  2015, \mn@doi [\apj] {10.1088/0004-637X/810/2/160}, \href {https://ui.adsabs.harvard.edu/abs/2015ApJ...810..160G} {810, 160}
    
    \bibitem[\protect\citeauthoryear{{Ghirlanda} et~al.,}{{Ghirlanda} et~al.}{2019}]{Ghirlanda19Compact}
    {Ghirlanda} G.,  et~al., 2019, \mn@doi [Science] {10.1126/science.aau8815}, \href {https://ui.adsabs.harvard.edu/abs/2019Sci...363..968G} {363, 968}
    
    \bibitem[\protect\citeauthoryear{{Gill} \& {Granot}}{{Gill} \& {Granot}}{2018}]{Gill18Afterglow}
    {Gill} R.,  {Granot} J.,  2018, \mn@doi [\mnras] {10.1093/mnras/sty1214}, \href {https://ui.adsabs.harvard.edu/abs/2018MNRAS.478.4128G} {478, 4128}
    
    \bibitem[\protect\citeauthoryear{{Gill} \& {Granot}}{{Gill} \& {Granot}}{2023}]{Gill23GRB}
    {Gill} R.,  {Granot} J.,  2023, \mn@doi [\mnras] {10.1093/mnrasl/slad075}, \href {https://ui.adsabs.harvard.edu/abs/2023MNRAS.524L..78G} {524, L78}
    
    \bibitem[\protect\citeauthoryear{{Gottlieb}, {Lalakos}, {Bromberg}, {Liska}  \& {Tchekhovskoy}}{{Gottlieb} et~al.}{2022}]{Gottlieb223DGRMHD}
    {Gottlieb} O.,  {Lalakos} A.,  {Bromberg} O.,  {Liska} M.,   {Tchekhovskoy} A.,  2022, \mn@doi [\mnras] {10.1093/mnras/stab3784}, \href {https://ui.adsabs.harvard.edu/abs/2022MNRAS.510.4962G} {510, 4962}
    
    \bibitem[\protect\citeauthoryear{{Granot}}{{Granot}}{2005}]{Granot05Afterglow}
    {Granot} J.,  2005, \mn@doi [\apj] {10.1086/432676}, \href {https://ui.adsabs.harvard.edu/abs/2005ApJ...631.1022G} {631, 1022}
    
    \bibitem[\protect\citeauthoryear{{Granot} \& {Kumar}}{{Granot} \& {Kumar}}{2003}]{Granot03Constraining}
    {Granot} J.,  {Kumar} P.,  2003, \mn@doi [\apj] {10.1086/375489}, \href {https://ui.adsabs.harvard.edu/abs/2003ApJ...591.1086G} {591, 1086}
    
    \bibitem[\protect\citeauthoryear{{Granot} \& {Sari}}{{Granot} \& {Sari}}{2002}]{granot02shape}
    {Granot} J.,  {Sari} R.,  2002, \mn@doi [\apj] {10.1086/338966}, \href {https://ui.adsabs.harvard.edu/abs/2002ApJ...568..820G} {568, 820}
    
    \bibitem[\protect\citeauthoryear{{Granot}, {Panaitescu}, {Kumar}  \& {Woosley}}{{Granot} et~al.}{2002}]{granot02offaxis}
    {Granot} J.,  {Panaitescu} A.,  {Kumar} P.,   {Woosley} S.~E.,  2002, \mn@doi [\apjl] {10.1086/340991}, \href {https://ui.adsabs.harvard.edu/abs/2002ApJ...570L..61G} {570, L61}
    
    \bibitem[\protect\citeauthoryear{{Granot}, {Ramirez-Ruiz}  \& {Perna}}{{Granot} et~al.}{2005}]{Granot05AfterglowObservations}
    {Granot} J.,  {Ramirez-Ruiz} E.,   {Perna} R.,  2005, \mn@doi [\apj] {10.1086/431477}, \href {https://ui.adsabs.harvard.edu/abs/2005ApJ...630.1003G} {630, 1003}
    
    \bibitem[\protect\citeauthoryear{{Granot}, {Guetta}  \& {Gill}}{{Granot} et~al.}{2017}]{Granot17Lessons}
    {Granot} J.,  {Guetta} D.,   {Gill} R.,  2017, \mn@doi [\apjl] {10.3847/2041-8213/aa991d}, \href {https://ui.adsabs.harvard.edu/abs/2017ApJ...850L..24G} {850, L24}
    
    \bibitem[\protect\citeauthoryear{{Granot}, {De Colle}  \& {Ramirez-Ruiz}}{{Granot} et~al.}{2018}]{Granot18Off}
    {Granot} J.,  {De Colle} F.,   {Ramirez-Ruiz} E.,  2018, \mn@doi [\mnras] {10.1093/mnras/sty2454}, \href {https://ui.adsabs.harvard.edu/abs/2018MNRAS.481.2711G} {481, 2711}
    
    \bibitem[\protect\citeauthoryear{{Hajela} et~al.,}{{Hajela} et~al.}{2019}]{Hajela19Two}
    {Hajela} A.,  et~al., 2019, \mn@doi [\apjl] {10.3847/2041-8213/ab5226}, \href {https://ui.adsabs.harvard.edu/abs/2019ApJ...886L..17H} {886, L17}
    
    \bibitem[\protect\citeauthoryear{{Huang}, {Wu}, {Dai}, {Ma}  \& {Lu}}{{Huang} et~al.}{2004}]{Huang04Rebrightening}
    {Huang} Y.~F.,  {Wu} X.~F.,  {Dai} Z.~G.,  {Ma} H.~T.,   {Lu} T.,  2004, \mn@doi [\apj] {10.1086/382202}, \href {https://ui.adsabs.harvard.edu/abs/2004ApJ...605..300H} {605, 300}
    
    \bibitem[\protect\citeauthoryear{{Kathirgamaraju}, {Barniol Duran}  \& {Giannios}}{{Kathirgamaraju} et~al.}{2018}]{Kathirgamaraju18Off}
    {Kathirgamaraju} A.,  {Barniol Duran} R.,   {Giannios} D.,  2018, \mn@doi [\mnras] {10.1093/mnrasl/slx175}, \href {https://ui.adsabs.harvard.edu/abs/2018MNRAS.473L.121K} {473, L121}
    
    \bibitem[\protect\citeauthoryear{{Keinan} \& {Arcavi}}{{Keinan} \& {Arcavi}}{2024}]{Keinan24Coordinated}
    {Keinan} I.,  {Arcavi} I.,  2024, \mn@doi [arXiv e-prints] {10.48550/arXiv.2405.17558}, \href {https://ui.adsabs.harvard.edu/abs/2024arXiv240517558K} {p. arXiv:2405.17558}
    
    \bibitem[\protect\citeauthoryear{{Kobayashi}}{{Kobayashi}}{2000}]{Kobayashi00Light}
    {Kobayashi} S.,  2000, \mn@doi [\apj] {10.1086/317869}, \href {https://ui.adsabs.harvard.edu/abs/2000ApJ...545..807K} {545, 807}
    
    \bibitem[\protect\citeauthoryear{{Kobayashi} \& {Sari}}{{Kobayashi} \& {Sari}}{2000}]{kobayashi00optical}
    {Kobayashi} S.,  {Sari} R.,  2000, \mn@doi [\apj] {10.1086/317021}, \href {https://ui.adsabs.harvard.edu/abs/2000ApJ...542..819K} {542, 819}
    
    \bibitem[\protect\citeauthoryear{Kobayashi, Piran  et~al.}{Kobayashi et~al.}{1999}]{kobayashi1999hydrodynamics}
    Kobayashi S.,  Piran T.,   et~al., 1999, The Astrophysical Journal, 513, 669
    
    \bibitem[\protect\citeauthoryear{{Komesh}, {Grossan}, {Maksut}, {Abdikamalov}, {Krugov}  \& {Smoot}}{{Komesh} et~al.}{2023}]{Komesh23Evolution}
    {Komesh} T.,  {Grossan} B.,  {Maksut} Z.,  {Abdikamalov} E.,  {Krugov} M.,   {Smoot} G.~F.,  2023, \mn@doi [\mnras] {10.1093/mnras/stad538}, \href {https://ui.adsabs.harvard.edu/abs/2023MNRAS.520.6104K} {520, 6104}
    
    \bibitem[\protect\citeauthoryear{{Kumar} \& {Granot}}{{Kumar} \& {Granot}}{2003}]{Kumar03Evolution}
    {Kumar} P.,  {Granot} J.,  2003, \mn@doi [\apj] {10.1086/375186}, \href {https://ui.adsabs.harvard.edu/abs/2003ApJ...591.1075K} {591, 1075}
    
    \bibitem[\protect\citeauthoryear{{Lamb} \& {Kobayashi}}{{Lamb} \& {Kobayashi}}{2017}]{Lamb17Electromagnetic}
    {Lamb} G.~P.,  {Kobayashi} S.,  2017, \mn@doi [\mnras] {10.1093/mnras/stx2345}, \href {https://ui.adsabs.harvard.edu/abs/2017MNRAS.472.4953L} {472, 4953}
    
    \bibitem[\protect\citeauthoryear{{Lamb} \& {Kobayashi}}{{Lamb} \& {Kobayashi}}{2018}]{Lamb18GRB}
    {Lamb} G.~P.,  {Kobayashi} S.,  2018, \mn@doi [\mnras] {10.1093/mnras/sty1108}, \href {https://ui.adsabs.harvard.edu/abs/2018MNRAS.478..733L} {478, 733}
    
    \bibitem[\protect\citeauthoryear{{Lamb} \& {Kobayashi}}{{Lamb} \& {Kobayashi}}{2019}]{Lamb19Reverse}
    {Lamb} G.~P.,  {Kobayashi} S.,  2019, \mn@doi [\mnras] {10.1093/mnras/stz2252}, \href {https://ui.adsabs.harvard.edu/abs/2019MNRAS.489.1820L} {489, 1820}
    
    \bibitem[\protect\citeauthoryear{{Lamb} et~al.,}{{Lamb} et~al.}{2021a}]{Lamb21Inclination}
    {Lamb} G.~P.,  et~al., 2021a, \mn@doi [Universe] {10.3390/universe7090329}, \href {https://ui.adsabs.harvard.edu/abs/2021Univ....7..329L} {7, 329}
    
    \bibitem[\protect\citeauthoryear{{Lamb}, {Kann}, {Fern{\'a}ndez}, {Mandel}, {Levan}  \& {Tanvir}}{{Lamb} et~al.}{2021b}]{Lamb21GRB}
    {Lamb} G.~P.,  {Kann} D.~A.,  {Fern{\'a}ndez} J.~J.,  {Mandel} I.,  {Levan} A.~J.,   {Tanvir} N.~R.,  2021b, \mn@doi [\mnras] {10.1093/mnras/stab2071}, \href {https://ui.adsabs.harvard.edu/abs/2021MNRAS.506.4163L} {506, 4163}
    
    \bibitem[\protect\citeauthoryear{{Laskar} et~al.,}{{Laskar} et~al.}{2013}]{Laskar13Reverse}
    {Laskar} T.,  et~al., 2013, \mn@doi [\apj] {10.1088/0004-637X/776/2/119}, \href {https://ui.adsabs.harvard.edu/abs/2013ApJ...776..119L} {776, 119}
    
    \bibitem[\protect\citeauthoryear{{Laskar} et~al.,}{{Laskar} et~al.}{2019}]{Laskar19Reverse}
    {Laskar} T.,  et~al., 2019, \mn@doi [\apj] {10.3847/1538-4357/ab40ce}, \href {https://ui.adsabs.harvard.edu/abs/2019ApJ...884..121L} {884, 121}
    
    \bibitem[\protect\citeauthoryear{{Laskar} et~al.,}{{Laskar} et~al.}{2023}]{Laskar23Radio}
    {Laskar} T.,  et~al., 2023, \mn@doi [\apjl] {10.3847/2041-8213/acbfad}, \href {https://ui.adsabs.harvard.edu/abs/2023ApJ...946L..23L} {946, L23}
    
    \bibitem[\protect\citeauthoryear{{Lazzati}, {Deich}, {Morsony}  \& {Workman}}{{Lazzati} et~al.}{2017}]{Lazzati17Off}
    {Lazzati} D.,  {Deich} A.,  {Morsony} B.~J.,   {Workman} J.~C.,  2017, \mn@doi [\mnras] {10.1093/mnras/stx1683}, \href {https://ui.adsabs.harvard.edu/abs/2017MNRAS.471.1652L} {471, 1652}
    
    \bibitem[\protect\citeauthoryear{{Lazzati}, {Perna}, {Morsony}, {Lopez-Camara}, {Cantiello}, {Ciolfi}, {Giacomazzo}  \& {Workman}}{{Lazzati} et~al.}{2018}]{Lazzati18Late}
    {Lazzati} D.,  {Perna} R.,  {Morsony} B.~J.,  {Lopez-Camara} D.,  {Cantiello} M.,  {Ciolfi} R.,  {Giacomazzo} B.,   {Workman} J.~C.,  2018, \mn@doi [\prl] {10.1103/PhysRevLett.120.241103}, \href {https://ui.adsabs.harvard.edu/abs/2018PhRvL.120x1103L} {120, 241103}
    
    \bibitem[\protect\citeauthoryear{{Li} et~al.,}{{Li} et~al.}{2012}]{Li12Comprehensive}
    {Li} L.,  et~al., 2012, \mn@doi [\apj] {10.1088/0004-637X/758/1/27}, \href {https://ui.adsabs.harvard.edu/abs/2012ApJ...758...27L} {758, 27}
    
    \bibitem[\protect\citeauthoryear{{Li} et~al.,}{{Li} et~al.}{2024}]{Li24Nature}
    {Li} M.~L.,  et~al., 2024, \mn@doi [arXiv e-prints] {10.48550/arXiv.2411.07973}, \href {https://ui.adsabs.harvard.edu/abs/2024arXiv241107973L} {p. arXiv:2411.07973}
    
    \bibitem[\protect\citeauthoryear{{Margutti} et~al.,}{{Margutti} et~al.}{2018}]{Margutti18Binary}
    {Margutti} R.,  et~al., 2018, \mn@doi [\apjl] {10.3847/2041-8213/aab2ad}, \href {https://ui.adsabs.harvard.edu/abs/2018ApJ...856L..18M} {856, L18}
    
    \bibitem[\protect\citeauthoryear{{M{\'e}sz{\'a}ros} \& {Rees}}{{M{\'e}sz{\'a}ros} \& {Rees}}{1997}]{Meszaros97Optical}
    {M{\'e}sz{\'a}ros} P.,  {Rees} M.~J.,  1997, \mn@doi [\apj] {10.1086/303625}, \href {https://ui.adsabs.harvard.edu/abs/1997ApJ...476..232M} {476, 232}
    
    \bibitem[\protect\citeauthoryear{{Mooley} et~al.,}{{Mooley} et~al.}{2018}]{Mooley18Superluminal}
    {Mooley} K.~P.,  et~al., 2018, \mn@doi [\nat] {10.1038/s41586-018-0486-3}, \href {https://ui.adsabs.harvard.edu/abs/2018Natur.561..355M} {561, 355}
    
    \bibitem[\protect\citeauthoryear{{Nakar}, {Piran}  \& {Granot}}{{Nakar} et~al.}{2002}]{Nakar02Detectability}
    {Nakar} E.,  {Piran} T.,   {Granot} J.,  2002, \mn@doi [\apj] {10.1086/342791}, \href {https://ui.adsabs.harvard.edu/abs/2002ApJ...579..699N} {579, 699}
    
    \bibitem[\protect\citeauthoryear{{Panaitescu}}{{Panaitescu}}{2007}]{Panaitescu07Jet}
    {Panaitescu} A.,  2007, \mn@doi [\mnras] {10.1111/j.1365-2966.2007.12084.x}, \href {https://ui.adsabs.harvard.edu/abs/2007MNRAS.380..374P} {380, 374}
    
    \bibitem[\protect\citeauthoryear{{Panaitescu} \& {Kumar}}{{Panaitescu} \& {Kumar}}{2003}]{Panaitescu03Effect}
    {Panaitescu} A.,  {Kumar} P.,  2003, \mn@doi [\apj] {10.1086/375563}, \href {https://ui.adsabs.harvard.edu/abs/2003ApJ...592..390P} {592, 390}
    
    \bibitem[\protect\citeauthoryear{{Pang} \& {Dai}}{{Pang} \& {Dai}}{2024a}]{Pang24Reverse_1}
    {Pang} S.-L.,  {Dai} Z.-G.,  2024a, \mn@doi [\mnras] {10.1093/mnras/stae197}, \href {https://ui.adsabs.harvard.edu/abs/2024MNRAS.528.2066P} {528, 2066}
    
    \bibitem[\protect\citeauthoryear{{Pang} \& {Dai}}{{Pang} \& {Dai}}{2024b}]{Pang24Reverse_2}
    {Pang} S.-L.,  {Dai} Z.-G.,  2024b, \mn@doi [\apj] {10.3847/1538-4357/ad9007}, \href {https://ui.adsabs.harvard.edu/abs/2024ApJ...977..123P} {977, 123}
    
    \bibitem[\protect\citeauthoryear{{Pe'er} \& {Wijers}}{{Pe'er} \& {Wijers}}{2006}]{Peer06Signature}
    {Pe'er} A.,  {Wijers} R. A.~M.~J.,  2006, \mn@doi [\apj] {10.1086/500969}, \href {https://ui.adsabs.harvard.edu/abs/2006ApJ...643.1036P} {643, 1036}
    
    \bibitem[\protect\citeauthoryear{{Perley} et~al.,}{{Perley} et~al.}{2014}]{Perley14Afterglow}
    {Perley} D.~A.,  et~al., 2014, \mn@doi [\apj] {10.1088/0004-637X/781/1/37}, \href {https://ui.adsabs.harvard.edu/abs/2014ApJ...781...37P} {781, 37}
    
    \bibitem[\protect\citeauthoryear{{Piran}, {Shemi}  \& {Narayan}}{{Piran} et~al.}{1993}]{Piran93Hydrodynamics}
    {Piran} T.,  {Shemi} A.,   {Narayan} R.,  1993, \mn@doi [\mnras] {10.1093/mnras/263.4.861}, \href {https://ui.adsabs.harvard.edu/abs/1993MNRAS.263..861P} {263, 861}
    
    \bibitem[\protect\citeauthoryear{{Rhoads}}{{Rhoads}}{1997}]{Rhoads97How}
    {Rhoads} J.~E.,  1997, \mn@doi [\apjl] {10.1086/310876}, \href {https://ui.adsabs.harvard.edu/abs/1997ApJ...487L...1R} {487, L1}
    
    \bibitem[\protect\citeauthoryear{{Rossi}, {Lazzati}  \& {Rees}}{{Rossi} et~al.}{2002}]{Rossi02Afterglow}
    {Rossi} E.,  {Lazzati} D.,   {Rees} M.~J.,  2002, \mn@doi [\mnras] {10.1046/j.1365-8711.2002.05363.x}, \href {https://ui.adsabs.harvard.edu/abs/2002MNRAS.332..945R} {332, 945}
    
    \bibitem[\protect\citeauthoryear{{Rossi}, {Lazzati}, {Salmonson}  \& {Ghisellini}}{{Rossi} et~al.}{2004}]{Rossi04polarization}
    {Rossi} E.~M.,  {Lazzati} D.,  {Salmonson} J.~D.,   {Ghisellini} G.,  2004, \mn@doi [\mnras] {10.1111/j.1365-2966.2004.08165.x}, \href {https://ui.adsabs.harvard.edu/abs/2004MNRAS.354...86R} {354, 86}
    
    \bibitem[\protect\citeauthoryear{{Ryan}, {van Eerten}, {Piro}  \& {Troja}}{{Ryan} et~al.}{2020}]{Ryan20afterglowpy}
    {Ryan} G.,  {van Eerten} H.,  {Piro} L.,   {Troja} E.,  2020, \mn@doi [\apj] {10.3847/1538-4357/ab93cf}, \href {https://ui.adsabs.harvard.edu/abs/2020ApJ...896..166R} {896, 166}
    
    \bibitem[\protect\citeauthoryear{{Sadeh}}{{Sadeh}}{2024}]{Sadeh24Detecting}
    {Sadeh} I.,  2024, \mn@doi [\apj] {10.3847/1538-4357/ad3ba5}, \href {https://ui.adsabs.harvard.edu/abs/2024ApJ...967..170S} {967, 170}
    
    \bibitem[\protect\citeauthoryear{{Salafia} \& {Ghirlanda}}{{Salafia} \& {Ghirlanda}}{2022}]{Salafia22Structure}
    {Salafia} O.~S.,  {Ghirlanda} G.,  2022, \mn@doi [Galaxies] {10.3390/galaxies10050093}, \href {https://ui.adsabs.harvard.edu/abs/2022Galax..10...93S} {10, 93}
    
    \bibitem[\protect\citeauthoryear{{Sari}}{{Sari}}{1997}]{sari97hydrodynamics}
    {Sari} R.,  1997, \mn@doi [\apjl] {10.1086/310957}, \href {https://ui.adsabs.harvard.edu/abs/1997ApJ...489L..37S} {489, L37}
    
    \bibitem[\protect\citeauthoryear{{Sari} \& {M{\'e}sz{\'a}ros}}{{Sari} \& {M{\'e}sz{\'a}ros}}{2000}]{Sari00Impulsive}
    {Sari} R.,  {M{\'e}sz{\'a}ros} P.,  2000, \mn@doi [\apjl] {10.1086/312689}, \href {https://ui.adsabs.harvard.edu/abs/2000ApJ...535L..33S} {535, L33}
    
    \bibitem[\protect\citeauthoryear{{Sari} \& {Piran}}{{Sari} \& {Piran}}{1995}]{sari95hydrodynamics}
    {Sari} R.,  {Piran} T.,  1995, \mn@doi [\apjl] {10.1086/309835}, \href {https://ui.adsabs.harvard.edu/abs/1995ApJ...455L.143S} {455, L143}
    
    \bibitem[\protect\citeauthoryear{{Sari} \& {Piran}}{{Sari} \& {Piran}}{1999}]{Sari99Predictions}
    {Sari} R.,  {Piran} T.,  1999, \mn@doi [\apj] {10.1086/307508}, \href {https://ui.adsabs.harvard.edu/abs/1999ApJ...520..641S} {520, 641}
    
    \bibitem[\protect\citeauthoryear{{Sari}, {Piran}  \& {Narayan}}{{Sari} et~al.}{1998}]{sari98spectra}
    {Sari} R.,  {Piran} T.,   {Narayan} R.,  1998, \mn@doi [\apjl] {10.1086/311269}, \href {https://ui.adsabs.harvard.edu/abs/1998ApJ...497L..17S} {497, L17}
    
    \bibitem[\protect\citeauthoryear{{Sari}, {Piran}  \& {Halpern}}{{Sari} et~al.}{1999}]{Sari99Jets}
    {Sari} R.,  {Piran} T.,   {Halpern} J.~P.,  1999, \mn@doi [\apjl] {10.1086/312109}, \href {https://ui.adsabs.harvard.edu/abs/1999ApJ...519L..17S} {519, L17}
    
    \bibitem[\protect\citeauthoryear{{Tian}, {Qin}, {Du}, {Yi}  \& {Tang}}{{Tian} et~al.}{2022}]{Tian22Constraining}
    {Tian} X.,  {Qin} Y.,  {Du} M.,  {Yi} S.-X.,   {Tang} Y.-K.,  2022, \mn@doi [\apj] {10.3847/1538-4357/ac3de4}, \href {https://ui.adsabs.harvard.edu/abs/2022ApJ...925...54T} {925, 54}
    
    \bibitem[\protect\citeauthoryear{{Troja} et~al.,}{{Troja} et~al.}{2018}]{Troja18outflow}
    {Troja} E.,  et~al., 2018, \mn@doi [\mnras] {10.1093/mnrasl/sly061}, \href {https://ui.adsabs.harvard.edu/abs/2018MNRAS.478L..18T} {478, L18}
    
    \bibitem[\protect\citeauthoryear{{Vestrand} et~al.,}{{Vestrand} et~al.}{2014}]{Vestrand14Bright}
    {Vestrand} W.~T.,  et~al., 2014, \mn@doi [Science] {10.1126/science.1242316}, \href {https://ui.adsabs.harvard.edu/abs/2014Sci...343...38V} {343, 38}
    
    \bibitem[\protect\citeauthoryear{{Wang} et~al.,}{{Wang} et~al.}{2015}]{Wang15How}
    {Wang} X.-G.,  et~al., 2015, \mn@doi [\apjs] {10.1088/0067-0049/219/1/9}, \href {https://ui.adsabs.harvard.edu/abs/2015ApJS..219....9W} {219, 9}
    
    \bibitem[\protect\citeauthoryear{{Woods} \& {Loeb}}{{Woods} \& {Loeb}}{1999}]{Woods99Constraints}
    {Woods} E.,  {Loeb} A.,  1999, \mn@doi [\apj] {10.1086/307738}, \href {https://ui.adsabs.harvard.edu/abs/1999ApJ...523..187W} {523, 187}
    
    \bibitem[\protect\citeauthoryear{{Xie}, {Zrake}  \& {MacFadyen}}{{Xie} et~al.}{2018}]{Xie18Numerical}
    {Xie} X.,  {Zrake} J.,   {MacFadyen} A.,  2018, \mn@doi [\apj] {10.3847/1538-4357/aacf9c}, \href {https://ui.adsabs.harvard.edu/abs/2018ApJ...863...58X} {863, 58}
    
    \bibitem[\protect\citeauthoryear{{Xin} et~al.,}{{Xin} et~al.}{2023}]{Xin23Prompt}
    {Xin} L.,  et~al., 2023, \mn@doi [Nature Astronomy] {10.1038/s41550-023-01930-0}, \href {https://ui.adsabs.harvard.edu/abs/2023NatAs...7..724X} {7, 724}
    
    \bibitem[\protect\citeauthoryear{{Yamazaki}, {Ioka}  \& {Nakamura}}{{Yamazaki} et~al.}{2002}]{Yamazaki02Xray}
    {Yamazaki} R.,  {Ioka} K.,   {Nakamura} T.,  2002, \mn@doi [\apjl] {10.1086/341225}, \href {https://ui.adsabs.harvard.edu/abs/2002ApJ...571L..31Y} {571, L31}
    
    \bibitem[\protect\citeauthoryear{{Yi}, {Wu}  \& {Dai}}{{Yi} et~al.}{2013}]{Yi13early}
    {Yi} S.-X.,  {Wu} X.-F.,   {Dai} Z.-G.,  2013, \mn@doi [\apj] {10.1088/0004-637X/776/2/120}, \href {https://ui.adsabs.harvard.edu/abs/2013ApJ...776..120Y} {776, 120}
    
    \bibitem[\protect\citeauthoryear{{Zhang} \& {Kobayashi}}{{Zhang} \& {Kobayashi}}{2005}]{Zhang05Gamma}
    {Zhang} B.,  {Kobayashi} S.,  2005, \mn@doi [\apj] {10.1086/429787}, \href {https://ui.adsabs.harvard.edu/abs/2005ApJ...628..315Z} {628, 315}
    
    \bibitem[\protect\citeauthoryear{{Zhang}, {Fan}, {Dyks}, {Kobayashi}, {M{\'e}sz{\'a}ros}, {Burrows}, {Nousek}  \& {Gehrels}}{{Zhang} et~al.}{2006}]{Zhang06Physical}
    {Zhang} B.,  {Fan} Y.~Z.,  {Dyks} J.,  {Kobayashi} S.,  {M{\'e}sz{\'a}ros} P.,  {Burrows} D.~N.,  {Nousek} J.~A.,   {Gehrels} N.,  2006, \mn@doi [\apj] {10.1086/500723}, \href {https://ui.adsabs.harvard.edu/abs/2006ApJ...642..354Z} {642, 354}
    
    \bibitem[\protect\citeauthoryear{{Zhang} et~al.,}{{Zhang} et~al.}{2018}]{Zhang18Transition}
    {Zhang} B.~B.,  et~al., 2018, \mn@doi [Nature Astronomy] {10.1038/s41550-017-0309-8}, \href {https://ui.adsabs.harvard.edu/abs/2018NatAs...2...69Z} {2, 69}
    
    \bibitem[\protect\citeauthoryear{{Zhang}, {Liu}, {Geng}, {Wu}  \& {Wang}}{{Zhang} et~al.}{2022}]{Zhang22semi}
    {Zhang} Z.-L.,  {Liu} R.-Y.,  {Geng} J.-J.,  {Wu} X.-F.,   {Wang} X.-Y.,  2022, \mn@doi [\mnras] {10.1093/mnras/stac1198}, \href {https://ui.adsabs.harvard.edu/abs/2022MNRAS.513.4887Z} {513, 4887}
    
    \bibitem[\protect\citeauthoryear{{Zhang}, {Wang}  \& {Zheng}}{{Zhang} et~al.}{2024}]{Zhang24BOAT}
    {Zhang} B.,  {Wang} X.-Y.,   {Zheng} J.-H.,  2024, \mn@doi [Journal of High Energy Astrophysics] {10.1016/j.jheap.2024.01.002}, \href {https://ui.adsabs.harvard.edu/abs/2024JHEAp..41...42Z} {41, 42}
    
    \bibitem[\protect\citeauthoryear{{Zhao} et~al.,}{{Zhao} et~al.}{2020}]{Zhao20Statistical}
    {Zhao} W.,  et~al., 2020, \mn@doi [\apj] {10.3847/1538-4357/aba43a}, \href {https://ui.adsabs.harvard.edu/abs/2020ApJ...900..112Z} {900, 112}
    
    \bibitem[\protect\citeauthoryear{{van Eerten}, {van der Horst}  \& {MacFadyen}}{{van Eerten} et~al.}{2012}]{vanEerten12Gamma}
    {van Eerten} H.,  {van der Horst} A.,   {MacFadyen} A.,  2012, \mn@doi [\apj] {10.1088/0004-637X/749/1/44}, \href {https://ui.adsabs.harvard.edu/abs/2012ApJ...749...44V} {749, 44}
    
    \makeatother
    \end{thebibliography}



\appendix

\section{Supplemental material}

\begin{table}
\centering
\begin{threeparttable}
\renewcommand{\arraystretch}{1.5} 
\begin{tabular}{|c|c|c|c|c|}
\hline
\textbf{Phase} & \textbf{Shell Type} & \textbf{Shock Type} & \textbf{$g$} & \textbf{$b_\mathrm{p}$} \\
\hline
\multirow{4}{*}{$t<t_\Delta$} 
  & \multirow{2}{*}{Thin} & FS & $0$ & $b$ \\
  &                       & RS & $0$ & $b$ \\
  \cline{2-5}
  & \multirow{2}{*}{Thick\tnote{a}} & FS & $\frac{2-k}{2(4-k)}$ & $\frac{a}{2(4-k)}$ \\
  &                        & RS & $\frac{2-k}{2(4-k)}$ & $\frac{a}{2(4-k)}$ \\
\hline
\multirow{4}{*}{$t>t_\Delta$} 
  & \multirow{2}{*}{Thin} & FS & $\frac{3-k}{2(4-k)}$ & $\frac{a}{2(4-k)}$ \\
  &                       & RS & $\frac{7-2k}{4(4-k)}$ & $\frac{a(7-2k)+2b(k-4)}{4(k^2-7k+12)}$ \\
  \cline{2-5}
  & \multirow{2}{*}{Thick} & FS & $\frac{3-k}{2(4-k)}$ & $\frac{a}{2(4-k)}$ \\
  &                        & RS & $\frac{7-2k}{4(4-k)}$ & $\frac{a}{2(4-k)}$ \\
\hline
\end{tabular}
\caption{Values of $g$ and $b_\mathrm{p}$ for thin and thick shells. FS refers to the forward shock, RS refers to the reverse shock, and $t_\Delta$ denotes the shock crossing time. The values of $g$ are taken from \citet{Yi13early}. The values of $b_\mathrm{p}$ for thick shells come from Eq.~(\ref{eq:gamma_delta}). For thin shells, it is obtained using relation $\Theta^{-b_\mathrm{p}} \propto \Gamma_0(\theta) t_\Delta(\theta)^g$.}
\label{tab:g_bp_values}
\begin{tablenotes}
\footnotesize
\item[a] For thick shells, the pre–shock crossing values of $g$ and $b_\mathrm{p}$ are valid only after the RS becomes relativistic, i.e., for $t_\mathrm{N}<t<t_\Delta$, where $t_\mathrm{N}$ is given by Eq.~(\ref{eq:tN}).
\end{tablenotes}
\end{threeparttable}
\end{table}

\begin{table}
\centering
\setlength{\tabcolsep}{4pt}
\begin{tabular}{|c|c|}
\hline
\textbf{Shell type} & \textbf{Expression} \\
\hline
\multirow{3}{*}{\raisebox{-8.5ex}{\begin{tabular}{c} FS, \\ thick shell \end{tabular}}} 
 & 
 $ \nu_\mathrm{m} \propto E^{\frac{1}{2}} \begin{cases} 
 (t/t_\Delta)^{-1}, & t < t_\Delta \\ 
 (t/t_\Delta)^{-\frac{3}{2}}, & t > t_\Delta 
 \end{cases}$ \\
\cline{2-2}
 & 
 $\nu_\mathrm{c} \propto E^{\frac{3k-4}{2(4-k)}} \begin{cases} 
 (t/t_\Delta)^{\frac{3k-4}{4-k}}, & t < t_\Delta \\ 
 (t/t_\Delta)^{\frac{3k-4}{2(4-k)}}, & t > t_\Delta 
 \end{cases}$ \\
\cline{2-2}
 & 
 $F_{\nu, \text{max}} \propto E^{\frac{8-3k}{2(4-k)}} \begin{cases} 
 (t/t_\Delta)^{\frac{4-2k}{4-k}}, & t < t_\Delta \\ 
 (t/t_\Delta)^{\frac{-k}{2(4-k)}}, & t > t_\Delta 
 \end{cases}$ \\
\hline
\multirow{3}{*}{\raisebox{-8.5ex}{\begin{tabular}{c} FS, \\ thin shell \end{tabular}}} 
 & 
 $ \nu_\mathrm{m} \propto \Gamma_0^{\frac{4-3k}{3-k}} E^{\frac{-k}{2(3-k)}} \begin{cases} 
 (t/t_\Delta)^{-\frac{k}{2}}, & t < t_\Delta \\ 
 (t/t_\Delta)^{-\frac{3}{2}}, & t > t_\Delta 
 \end{cases}$ \\
\cline{2-2}
 & 
 $\nu_\mathrm{c} \propto \Gamma_0^{\frac{4-3k}{3-k}} E^{\frac{3k-4}{2(3-k)}} \begin{cases} 
 (t/t_\Delta)^{\frac{3k-4}{2}}, & t < t_\Delta \\ 
 (t/t_\Delta)^{\frac{3k-4}{2(4-k)}}, & t > t_\Delta
 \end{cases}$ \\
\cline{2-2}
 & 
 $F_{\nu, \text{max}} \propto \Gamma_0^{\frac{k}{3-k}} E^{\frac{3(2-k)}{2(3-k)}} \begin{cases} 
 (t/t_\Delta)^{\frac{6-3k}{2}}, & t < t_\Delta \\ 
 (t/t_\Delta)^{\frac{-k}{2(4-k)}}, & t > t_\Delta
 \end{cases}$ \\
\hline
\multirow{3}{*}{\raisebox{-8.5ex}{\begin{tabular}{c} RS, \\ thick shell \end{tabular}}} 
 & 
 $\nu_\mathrm{m} \propto \Gamma_0^2 E^{\frac{-k}{2(4-k)}} \begin{cases} 
 (t/t_\Delta)^{\frac{-k}{4-k}}, & t < t_\Delta \\ 
 (t/t_\Delta)^{\frac{14k-73}{12(4-k)}}, & t > t_\Delta 
 \end{cases}$ \\
\cline{2-2}
 & 
 $\nu_\mathrm{c} \propto E^{\frac{3k-4}{2(4-k)}} \begin{cases} 
 (t/t_\Delta)^{\frac{3k-4}{4-k}}, & t < t_\Delta \\ 
 (t/t_\Delta)^{\frac{14k-73}{12(4-k)}}, & t > t_\Delta 
 \end{cases}$ \\
\cline{2-2}
 & 
 $F_{\nu, \text{max}} \propto \Gamma_0^{-1} E^{\frac{10-3k}{2(4-k)}} \begin{cases} 
 (t/t_\Delta)^{\frac{2-k}{4-k}}, & t < t_\Delta \\ 
 (t/t_\Delta)^{\frac{10k-47}{12(4-k)}}, & t > t_\Delta 
 \end{cases}$ \\
\hline
\multirow{3}{*}{\raisebox{-8.5ex}{\begin{tabular}{c} RS, \\ thin shell \end{tabular}}}
 & 
 $\nu_\mathrm{m} \propto \Gamma_0^{\frac{6-k}{3-k}} E^{\frac{-k}{2(3-k)}} \begin{cases} 
 (t/t_\Delta)^{\frac{12-5k}{2}}, & t < t_\Delta \\ 
 (t/t_\Delta)^{\frac{14k-73}{12(4-k)}}, & t > t_\Delta 
 \end{cases}$ \\
\cline{2-2}
 & 
 $\nu_\mathrm{c} \propto \Gamma_0^{\frac{4-3k}{3-k}} E^{\frac{3k-4}{2(3-k)}} \begin{cases} 
 (t/t_\Delta)^{\frac{3k-4}{2}}, & t < t_\Delta \\ 
 (t/t_\Delta)^{\frac{14k-73}{12(4-k)}}, & t > t_\Delta 
 \end{cases}$ \\
\cline{2-2}
 & 
 $F_{\nu, \text{max}} \propto \Gamma_0^{\frac{3}{3-k}} E^{\frac{3(2-k)}{2(3-k)}} \begin{cases} 
 (t/t_\Delta)^{\frac{3-2k}{2}}, & t < t_\Delta \\ 
 (t/t_\Delta)^{\frac{10k-47}{12(4-k)}}, & t > t_\Delta 
 \end{cases}$ \\
\hline
\end{tabular}
\caption{The dependence of $\nu_\mathrm{m}$, $\nu_\mathrm{c}$, and $F_{\nu,\text{max}}$ on the energy $E$, initial Lorentz factor $\Gamma_0$, shock crossing time $t_\Delta$, and time $t$ for both reverse shock and forward shock. These expressions were derived by \citet{Yi13early} and are provided here for completeness. The dependence on $a$ and $b$ arise because, for a given $\theta$, $E$ and $\Gamma_0$ are functions of these parameters (cf. Eq.~\ref{eq:jet_structure}). For thin shells, $t_\Delta \propto E^{1/(3-k)}\Gamma_0^{(8-2k)/(3-k)}$. Substituting this into the expressions for $\nu_\mathrm{m}$, $\nu_\mathrm{c}$, and $F_{\nu,\text{max}}$, we obtain results that match the corresponding expressions for a thick-shell FS after shock crossing, which are independent of the initial Lorentz factor $\Gamma_0$.} 
\label{tab:dep_on_E_LF}
\end{table}

\begin{table*}
\centering
\begin{tabular}{|l|l|l|}
\hline
\textbf{Shock Type} & \textbf{Phase} & \textbf{Temporal slope} \\ \hline
\multirow{2}{*}{\raisebox{-2.0ex}{FS}} 
& $t < t_\Delta$ (thick shell) & \rule{0pt}{4ex}$\frac{2(k-2)}{a}-\frac{1}{4}k(p+5)+3 \quad \left[  \frac{2(k-2)}{a}-\frac{1}{4}k(p+2)+2 \right]$ \rule[-2ex]{0pt}{0pt} \\ \cline{2-3}
& $t > t_\Delta$ (thick and thin shells) & \rule{0pt}{4ex}$\frac{2(k-3)}{a}-\frac{1}{4}k(p+5)+3 \quad \left[ \frac{2(k-3)}{a}-\frac{1}{4}k(p+2)+2 \right]$ \rule[-2ex]{0pt}{0pt} \\ \hline
\multirow{3}{*}{\raisebox{-6.0ex}{RS}} 
& $t < t_\Delta$ (thick shell) & \rule{0pt}{4ex}$-\frac{(k-2)(b(p-2)-2)}{a}-\frac{1}{4}k(p+5)+3$\rule[-2ex]{0pt}{0pt} \\ \cline{2-3}
& $t > t_\Delta$ (thick shell) & \rule{0pt}{4ex}$\frac{k(-6k(p+5)+7p+219)+73p-399}{24(k-4)}-\frac{(2k-7)(b(p-2)-2)}{2a}$\rule[-2ex]{0pt}{0pt} \\ \cline{2-3}
& $t > t_\Delta$ (thin shell) & \rule{0pt}{4ex}$\frac{-3a(2k-7)(k(p+5)-12)+2b\left(8k^2-58k+93\right)p+6(2k-7)(2(b+2)k-3(b+4))}{12(a(2k-7)-2b(k-4))}$\rule[-2ex]{0pt}{0pt} \\ \hline
\end{tabular}
\caption{Temporal slopes of de-beamed emission for thin and thick shells in the regime $\nu_\mathrm{m} < \nu < \nu_\mathrm{c}$. To obtain expressions for $ \nu < \nu_\mathrm{m}$ regime, substitute $p$ with $1/3$. The expressions in the square brackets are for regime $\nu > \nu_\mathrm{c}$. Here, $t_\Delta$ denotes the shock crossing time.}
\label{tab:temporal_indices}
\end{table*}

\bsp	
\label{lastpage}
\end{document}